\documentclass[examplefnt,biber]{nowfnt} 

\usepackage[utf8]{inputenc}
\usepackage{multirow}
\usepackage{graphicx}
\usepackage{amsmath}
\usepackage[psamsfonts]{amssymb}
\usepackage{amsxtra}
\usepackage{threeparttable}
\usepackage{caption}
\usepackage{subcaption}
\usepackage{bbm}
\usepackage{booktabs}

\usepackage{tikz}
\usetikzlibrary{trees,positioning,shapes,shadows,arrows}

\title{A Review of Speech-centric Trustworthy Machine Learning: Privacy, Safety, and Fairness}

{%
 }

\usepackage{mwe}

\author{Tiantian Feng}
\author{Rajat Hebbar$^{\dagger}$}
\author{Nicholas Mehlman$^{\dagger}$}
\author{Xuan Shi$^{\dagger}$}
\author{Aditya Kommineni$^{\dagger}$}
\author{Shrikanth Narayanan}

\affil[1]{University of Southern California, Los Angeles, USA; \\
email: tiantiaf@usc.edu}

\articledatabox{$^{\dagger}$: Contribute equally in terms of text.}

\begin{document}

\makeabstracttitle

\begin{abstract}
Speech-centric machine learning systems have revolutionized a number of leading industries ranging from transportation and healthcare to education and defense, fundamentally reshaping how people live, work, and interact with each other. However, recent studies have demonstrated that many speech-centric ML systems may need to be considered more trustworthy for broader deployment. Specifically, concerns over privacy breaches, discriminating performance, and vulnerability to adversarial attacks have all been discovered in ML research fields. In order to address the above challenges and risks, a significant number of efforts have been made to ensure these ML systems are trustworthy, especially private, safe, and fair. In this paper, we conduct the first comprehensive survey on speech-centric trustworthy ML topics related to privacy, safety, and fairness. In addition to serving as a summary report for the research community, we highlight several promising future research directions to inspire researchers who wish to explore further in this area.
\end{abstract}

\chapter{Introduction}

In the last few years, machine learning (ML), particularly deep learning, has empowered tremendous breakthroughs in a variety of research fields and applications, including natural language processing (\cite{devlin2018bert}), image classification (\cite{he2016deep}), video recommendation (\cite{davidson2010youtube}), healthcare analysis (\cite{miotto2018deep}), and even mastering the chess game (\cite{silver2016mastering}). The deep learning model typically consists of multiple processing layers with a combination of both linear and non-linear computations. Although training a deep learning model with the multi-layer architecture demands the accumulation of massive datasets and access to large-scale computational infrastructures (\cite{bengio2021deep}), the trained model usually achieves state-of-the-art (SOTA) performance compared to the traditional modeling approaches. The broad success of deep learning has enabled a more profound understanding of the human condition (state, trait, behavior, interaction) and revolutionized technologies that support and enhance human experiences. Alongside the success that ML has had in these areas, significant progress has also been made in speech-centric ML.

Speech is a natural and prominent form of communication between humans that exists in almost every spectrum of human life, whether chatting with friends, discussing with colleagues, or having a remote call with the family. The advancement in speech-centric machine learning has enabled the ubiquitous usage of smart assistants such as Siri, Google Voice, and Alexa. In addition, speech-centric modeling has created numerous research topics in human behavior understanding, human-computer interface (HCI) (\cite{clark2019state}), and social media analysis, involving several widely used speech modeling techniques like automatic speech recognition (\cite{malik2021automatic}), speech emotion recognition (\cite{akccay2020speech}), automatic speaker verification (\cite{irum2019speaker}), and keyword spotting (\cite{warden2018speech}).

Despite the prospect of the broad deployment of ML systems in a wide range of speech-centric applications, two intertwined challenges remain unaddressed in most of these systems: understanding and illuminating the rich diversity across people and contexts while creating trustworthy ML technologies that work for everyone in all contexts. Trust is fundamental in human life, whether to trust friends, colleagues, family members, or AI-powered services. While ML practitioners, such as researchers, and decision-makers, conventionally focus on improving the system performance of ML models using performance metrics such as the F1 score, ensuring ML application that is trustworthy stays a challenging topic. In the past few years, we have witnessed a significant amount of research work targeting trustworthy AI and ML, and the objective of this paper is to provide a comprehensive review of related research activities, with an emphasis on speech-centric ML. This survey aims to outline salient design pillars and the latest research trends in speech-centric trustworthy ML.

Trustworthiness in ML has been defined differently across the literature. For example, \citet{huang2020survey} described the term trustworthiness based on the industry practices, involving the implementation of the certification process and explanation process. The certification process consists of testing and verification modules to detect potential fabrications or perturbations in the input data. The explanation refers to the ability to explain why ML reached a specific decision based on the input data. Furthermore, the ethics guidelines for trustworthy artificial intelligence published by EU (\cite{smuha2019eu}) recognized an AI system, to be considered trustworthy, must comply with laws and regulations, adhere to ethical principles, and function robustly. More recently, \citet{liu2022trustworthy} summarized the trustworthy AI from safety, fairness, explainability, privacy, accountability, and environmentally friendly aspects. Likewise, our review recognizes robustness, reliability, safety, security, inclusiveness, and equity as core design elements of the trustworthiness ML system. Based on these criteria, our paper surveys the literature on speech-centric trustworthy ML from privacy, safety, and fairness perspectives, as illustrated in Figure~\ref{fig:trustworthy_intro} \footnote{The figure in this paper uses images from https://openmoji.org/}:

\begin{figure*}[t]
    \begin{center}
        \includegraphics[width=\linewidth]{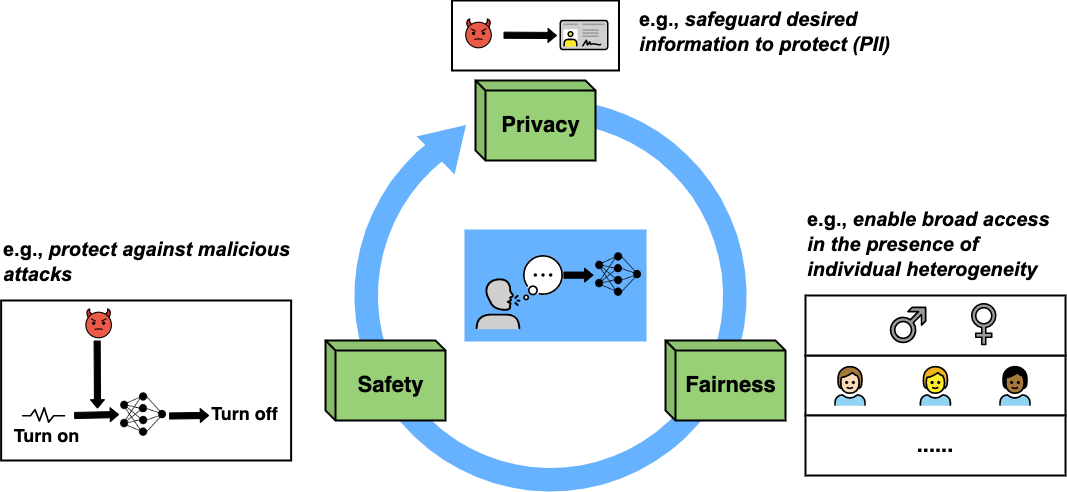}
    \end{center}
    \caption{Summary of key factors contribute to speech-centric trustworthy machine learning: privacy, safety, and fairness. } 
    \label{fig:trustworthy_intro}
\end{figure*}

\vspace{0.75mm}
\noindent \textbf{Privacy:} The speech-centric ML systems rely heavily on collecting speech data that is from, about, and for people in potentially sensitive environments and contexts, like homes, workplaces, hospitals, and schools. The collection of speech data frequently raises significant concerns about compromising user privacy, such as revealing sensitive information that people might want to keep private (\cite{liu2021machine}). It is critical to make sure that the speech data that is either shared by an individual or collected by an ML system is protected from unjustifiable and unauthorized uses.

\vspace{0.75mm}
\noindent \textbf{Safety:} Over the past few years, it has been discovered that ML systems are susceptible to adversarial attacks that aim to exploit vulnerabilities in the model's prediction function for malicious purposes (\cite{goodfellow2014explaining}). For example, by introducing minor perturbations to the speech data, a malicious actor can cause the keyword spotting model to drastically misclassify the desired input speech commands. Therefore, a trustworthy ML system must generate reliable outputs for the same inputs even if the inputs have been intentionally altered by the malicious attacker \citet{mehlman2023mel}.

\vspace{0.75mm}
\noindent \textbf{Fairness:} Recently, it has come to light that ML systems can perform unfairly. Why an ML system mistreats people is multifold (\cite{mehrabi2021survey}). One factor is the societal aspect, where the ML system generates biased outputs because of the societal biases in the training data or the assumptions/decisions throughout the ML development processes. Another reason that causes unfairness in AI is the imbalance of dataset characteristics, where limited data samples exist for some groups of people. As a result, the model needs to accommodate the needs of these groups to avoid biased outputs. It is also crucial to note that deploying unfair ML systems can amplify societal biases and data imbalance issues. To evaluate the trustworthiness of the speech-centric ML system, ML practitioners need to evaluate whether the ML model produces discriminatory outcomes towards individuals or groups.

The remainder of this article is organized as follows. Section 2 briefly summarizes popular speech-centric tasks, datasets, and SOTA modeling frameworks. Section 3 overviews the related survey papers in trustworthy speech-centric applications. Section 4 comprehensively discusses safety considerations in the speech-centric ML system. Section 5 discusses privacy risks and defenses in speech modeling. Section 6 reviews the emerging fairness issues with speech modeling tasks. Section 7 elaborates on potential developments and challenges in the future for speech-centric trustworthy machine learning. Finally, section 8 concludes with a summary of the key observations in this article. Specifically, our contributions are summarized as follows:

\begin{enumerate}
    \item To the best of our knowledge, this is the first review work to provide a comprehensive review of designing trustworthy ML focused on speech-centric modeling. We survey most, if not all, the published and pre-print works covering automatic speech recognition, speech emotion recognition, keyword spotting, and automatic speaker verification. 

    \item We create the taxonomies to systematically review design pillars related to the trustworthiness of speech-centric ML systems. We further compare a variety of literature on each key factor.

    \item We discuss the outstanding challenges of designing speech-centric ML systems facing trustworthiness considerations related to privacy, safety, and fairness. Based on the literature reviewed, we also discuss the challenges yet to be solved and suggest several promising future directions.

\end{enumerate}

\chapter{Speech-centric Machine Learning}
To begin with, we introduce the fundamentals of speech-centric ML systems to offer the audience a more in-depth understanding of the trustworthy aspect of such systems. Our aim is to provide a concise summary of speech-centric ML tasks, commonly used speech datasets, and speech modeling approaches.

\section{Speech-centric ML tasks}

\vspace{0.75mm}
\noindent \textbf{Automatic Speech Recognition (ASR)}: ASR system is one of the most prominent tasks in the speech domain, used to convert human speech into readable text. ASR techniques have found wide applications in modern human-computer interactive scenarios, such as Siri and Alexa. An ASR model typically involves pre-processing, feature extraction, classification, and the language model. The most widely adopted speech processing methods include framing, normalization, and pre-emphasis. After pre-processing, a feature extraction module extracts speech features as the input to the classifier. Mel-frequency cepstral coefficients (MFCCs) and Mel Spectrogram are commonly used speech features. The classification then predicts the spoken text based on the input features. Finally, a language model may be used to recognize the phoneme predicted by the classification model. It is worth noting that language models can significantly increase the efficiency of ASR systems, but it is not a necessary component in ASR systems. Many modern ASR systems can still function without using language models. Notably, word error rate (WER) is a standard metric to measure the performance of an ASR system, calculated as the number of errors in transcripts divided by the total number of spoken words. Readers interested in ASR systems may refer to \citet{malik2021automatic} for a systematic introduction.

\vspace{0.75mm}

\noindent \textbf{Speech Emotion Recognition (SER)}: The task of speech emotion recognition involves the classification of emotions (such as neutral, happy, sad, and angry) from speech signals, where the emotion labels are typically obtained from multiple human annotators. Equivalent to ASR, SER systems extract speech features passing through a classifier for emotion prediction. Due to the imbalanced label distributions in many existing SER datasets, the unweighted average recall (UAR) is often used as the evaluation metric for SER systems. For readers interested in a more comprehensive introduction to SER systems, we recommend referring to \citet{akccay2020speech}.

\vspace{0.75mm}
\noindent \textbf{Automatic speaker verification (ASV)}: ASV aims to determine the authorization of the claimed identity based on voice fingerprint. ASV is composed of two stages: the enrolment and the testing stage. During the enrolment stage, embeddings or latent representations are extracted from speakers and stored. In testing, embeddings from a test speaker are extracted and compared to the claimed speaker's embeddings. Verification pairs from the same speaker are referred to as \textit{genuine} pairs, while pairs from different speakers are "imposter" pairs (\cite{peri2022train}). Equal error rate (EER) is one of the most common metrics to evaluate the performance of ASV systems, which is the value at which the false acceptance rate (FAR) equals the false rejection rate (FRR). Further information on recent ASV developments can be found in \citet{irum2019speaker}.

\vspace{0.75mm}
\noindent \textbf{Keyword Spotting (KWS)}: Keyword spotting refers to the detection of predefined keywords or phrases in a speech recording, such as "Hello, Siri." This speech task is widely adopted in commercially available smart assistants such as Apple's Siri, Google Assistant, and Amazon's Alexa. Compared to ASR systems, KWS systems require substantially fewer computational resources and are well-suited for on-device learning.

\begin{table}
    \centering
    \caption{Table showing the List of speech datasets that can be used for training the ASR model, the SER model, the AVS model, and the KWS model.}
    \footnotesize
    \begin{tabular}{p{1.5cm}p{1.5cm}p{1.5cm}p{1.5cm}}

        \toprule
        \multicolumn{1}{c}{\textbf{Speech Task}} & 
        \multicolumn{1}{c}{\textbf{Dataset Name}} & 
        \multicolumn{1}{c}{\textbf{Year}} & 
        \multicolumn{1}{c}{\textbf{Hours}}\\

        \midrule

        & 
        \multicolumn{1}{l}{{Librispeech (\cite{panayotov2015librispeech})}} & 
        \multicolumn{1}{c}{{2015}} & 
        \multicolumn{1}{c}{{1,000}} \\

        & 
        \multicolumn{1}{l}{{WSJ (\cite{garofolo1993csr})}} & 
        \multicolumn{1}{c}{{1994}} & 
        \multicolumn{1}{c}{{162}} \\

        & 
        \multicolumn{1}{l}{{Common Voice (\cite{ardila2019common})}} & 
        \multicolumn{1}{c}{{2019}} & 
        \multicolumn{1}{c}{{1,900}} \\

        \multicolumn{1}{c}{{ASR}} & 
        \multicolumn{1}{l}{{The CHiME-5 (\cite{barker2018fifth})}} & 
        \multicolumn{1}{c}{{2018}} & 
        \multicolumn{1}{c}{{50}} \\

        & 
        \multicolumn{1}{l}{{TED-LIUM (\cite{rousseau2012ted})}} & 
        \multicolumn{1}{c}{{2012}} & 
        \multicolumn{1}{c}{{452}} \\

        & 
        \multicolumn{1}{l}{{The Spoken Wikipedia (\cite{kohn2016mining})}} & 
        \multicolumn{1}{c}{{2016}} & 
        \multicolumn{1}{c}{{1,005}} \\

        & 
        \multicolumn{1}{l}{{Voxpopuli (\cite{wang2021voxpopuli})}} & 
        \multicolumn{1}{c}{{2021}} & 
        \multicolumn{1}{c}{{1,800}} \\
        
        \midrule

        & 
        \multicolumn{1}{l}{{IEMOCAP (\cite{busso2008iemocap})}} & 
        \multicolumn{1}{c}{{2008}} & 
        \multicolumn{1}{c}{{12}} \\

        & 
        \multicolumn{1}{l}{{CMU-MOSEI (\cite{zadeh2018multimodal})}} & 
        \multicolumn{1}{c}{{2018}} & 
        \multicolumn{1}{c}{{65}} \\
        
        \multicolumn{1}{c}{{SER}} & 
        \multicolumn{1}{l}{{MSP-Podcast (\cite{martinez2020msp})}} & 
        \multicolumn{1}{c}{{2020}} & 
        \multicolumn{1}{c}{{100}} \\

        & 
        \multicolumn{1}{l}{{MELD (\cite{poria2018meld})}} & 
        \multicolumn{1}{c}{{2018}} & 
        \multicolumn{1}{c}{{-}} \\

        & 
        \multicolumn{1}{l}{{ESD (\cite{zhou2021seen})}} & 
        \multicolumn{1}{c}{{2021}} & 
        \multicolumn{1}{c}{{29}} \\

        \midrule
        & 
        \multicolumn{1}{l}{{Voxceleb1 (\cite{nagrani2017voxceleb})}} & 
        \multicolumn{1}{c}{{2017}} & 
        \multicolumn{1}{c}{{352}} \\

        \multicolumn{1}{c}{{AVS}} & 
        \multicolumn{1}{l}{{Voxceleb2 (\cite{chung2018voxceleb2})}} & 
        \multicolumn{1}{c}{{2018}} & 
        \multicolumn{1}{c}{{2,442}} \\

        & 
        \multicolumn{1}{l}{{VoxMovies (\cite{brown2021playing})}} & 
        \multicolumn{1}{c}{{2021}} & 
        \multicolumn{1}{c}{{-}} \\

        \midrule
        \multicolumn{1}{c}{{KWS}} & 
        \multicolumn{1}{l}{{Speech Commands (\cite{warden2018speech})}} & 
        \multicolumn{1}{c}{{2018}} & 
        \multicolumn{1}{c}{{-}} \\
        
        \bottomrule

    \end{tabular}
    
    \label{tab:speech_dataset}
\end{table}

\section{Speech Datasets}
The speech datasets are fundamental for training and testing speech-centric ML systems. This section provides a brief overview of the commonly used datasets for various downstream speech tasks. These datasets are also frequently used as benchmarks for evaluating speech-centric trustworthy ML. Then, we summarize the datasets along with their published years and total recording time in Table~\ref{tab:speech_dataset}.

\vspace{0.75mm}
\noindent \textbf{ASR}: Among all the datasets listed in Table~\ref{tab:speech_dataset}, Librispeech (\cite{panayotov2015librispeech}) is one of the most commonly used datasets for training and evaluating ASR algorithms. This dataset includes 1000 hours of audiobooks and transcriptions. Common Voice (\cite{ardila2019common}) is another widely used dataset supported by Mozilla. The dataset is entirely open source, and any people can contribute to the dataset by providing their speech recordings. Moreover, the CHiME-5 (\cite{barker2018fifth}) and TED-LIUM (\cite{rousseau2012ted}) datasets consist of audio recordings from the home environment and TED talks.

\vspace{0.75mm}
\noindent \textbf{SER}: One of the most frequently referred datasets in the SER field is the IEMOCAP dataset (\cite{busso2008iemocap}). This dataset consists of audio-visual data collected from 10 actors engaged in dyadic interactions. Each recorded utterance is provided with annotations into categorical emotion labels and transcripts. In addition to IEMOCAP, other commonly used datasets include the CMU-MOSEI dataset, which is based on YouTube videos (\cite{zadeh2018multimodal}), the MSP-Podcast dataset, which is based on podcasts (\cite{martinez2020msp}), and the MELD dataset, which is based on movies (\cite{poria2018meld}).

\vspace{0.75mm}
\noindent \textbf{AVS}: The two most frequently used datasets for training the AVS system are Voxceleb1 (\cite{nagrani2017voxceleb}) and Voxceleb2 (\cite{chung2018voxceleb2}). The voxceleb1 dataset contains speech data from over 1000 celebrities on Youtube. Shortly after the success of the voxceleb1 dataset, the authors introduced the voxceleb2 dataset which includes over 6000 celebrities' speech data on Youtube.

\vspace{0.75mm}
\noindent \textbf{KWS}: Surprisingly, there are few purposely designed datasets for this downstream task. The Google Speech Commands dataset (\cite{warden2018speech}) is the most commonly used KWS dataset that includes 35 frequently used spoken words from the everyday vocabulary. The dataset includes a total of 105,829 audio recordings from 2,618 speakers, wherein 2,112 speakers are in the training set and the rest are in the test set.

\section{Speech Modeling Approach}

\subsection{Conventional Modeling Approach}
Conventional speech modeling systems heavily relied on the Gaussian mixture model (GMM), Hidden Markov Model (HMM), and the support vector machine (SVM). For example, traditional speaker verification systems have used the Gaussian mixture model based universal background model (GMM-UBM) since early 2000 (\cite{reynolds2000speaker}). Later, the researchers proposed the i-vector framework (\cite{dehak2010front}), which reduced the high-dimensional GMM-UBM supervectors into low-dimensional vectors using factor analysis. The i-vector has been a SOTA technique in ASV systems for many years. In addition, HMM was one of the most widely used ASR modeling approaches before the deep learning approach gained popularity (\cite{trentin2001survey}). On the other hand, statistical descriptors of low-level speech features (e.g., pitch, intensity) were predominantly applied in emotion recognition and sentiment analysis tasks. OpenSMILE (\cite{eyben2010opensmile}) is one of the widely adopted tools that enable abundant research works using this approach. However, the performance of these traditional modeling frameworks is largely impacted by the channel variations and utterance variations of the input speech signals. 

\subsection{Deep Learning Approach}

In recent years, we have seen a wide variety of successes in applying deep learning to speech-centric systems. Deep speech (\cite{hannun2014deep}) is one of the most recognized deep neural models based on the recurrent neural network (RNN). This model has been a strong baseline in the ASR domain for years. Until a few years ago, with the popularity of transformer architecture (\cite{vaswani2017attention}), researchers have proposed the self-supervised learning framework like Wav2Vec 2.0 (\cite{baevski2020wav2vec}). Wav2Vec 2.0 has quickly become the SOTA machine learning model for ASR (\cite{baevski2020wav2vec}) and SER (\cite{chen2021exploring}). This model even achieves similar performance in the ASR task compared to humans. Around the same time, Google proposed a convolution-augmented transformer called Conformer (\cite{gulati2020conformer}) that reached competitive ASR performance to the Wav2Vec 2.0 model. More recently, OpenAI introduced its transformer-based ASR model called Whisper (\cite{radford2022robust}), which outperformed most existing works. Similar to the ASR task, many SOTA speaker verification systems are built upon the deep embedding, also known as x-vector, extracted from the deep neural network (\cite{snyder2018x}). 

\chapter{Related Surveys in Trustworthy Speech-centric Machine Learning}

In this section, we describe several related survey papers that cover the privacy and adversarial defense in speech applications. However, to the best of our knowledge, there are no survey papers focusing on fairness risks in speech-centric applications.

\section{Privacy}
\citet{cai2021generative} provides a detailed literature review on generative models, while the author briefly describes the use of generative speech models to protect user privacy. \citet{cai2021generative} introduces 2 usage cases, remote health monitoring, and voice assistance, based on Generative speech models. In both applications, the author presents GAN-based speech models that either obfuscate the sensitive attribute or transform the user voice into a common speech signal. \citet{cheng2022personal} is another review paper that focuses on privacy in personal voice assistant applications. This survey paper discusses voice privacy preservation mechanisms including encryption schemes, voice anonymization, and distributed learning. However, this paper does not provide comprehensive reviews on privacy attacks and also does not summarize the taxonomy of privacy-preserving methods. 

\section{Safety}

Apart from surveying privacy-related research topics, \citet{cheng2022personal} provides extensive reviews on security challenges in speech-centric applications. The author discusses a wide range of research works in mitigating adversarial attacks in ASR applications. This review covers popular adversarial attacks including PGD attacks, gradient decent attacks, etc. However, this review does not provide a categorization of adversarial attacks. Likewise, \citet{huynh2022adversarial} provides a survey on the safety aspect in speech-centric applications that categorize adversarial attacks based on types of attacking perturbations and threat models. Despite surveying different adversarial attacks in speech-centric applications, these survey papers do not discuss theories behind adversarial attacks and mitigations.

\section{Fairness}

Although many survey papers have discussed the fairness considerations in machine learning (e.g, \cite{mehrabi2021survey}), there is a lack of comprehensive literature studying bias and fairness in speech-centric applications. Most previous work in the speech domain focus on specific applications such as ASR, ASV and SER using traditional evaluation schemes.
For example, \citet{peri2022train} highlight the need for fairness-centric metrics that better highlights the bias in existing methods and evaluation schemes to improve fairness in ASV.
In this paper, we provide detailed reviews summarizing the recent works that target improving model fairness in ASR, ASV, and SER applications.

\chapter{Safety in Speech-centric Machine Learning}

Safety concerns in ML systems originated from adversarial attacks. Adversarial attacks involve a malicious actor (`adversary') who manipulates data samples with the intention of negatively affecting model performance. In a poisoning attack, the adversary manipulates data and/or model parameters during training, while in an evasion attack, they pass an adversarial sample to the model at evaluation time. In both scenarios, the adversary attempts to avoid detection by the model's users and/or administrators.

\section{Evasion Attacks and Defenses}

\subsection{Preliminaries}

Evasion attacks can be classified both by the adversary's knowledge of the targeted model (black box or white box) and by the attack's objective (targeted or untargeted). A black-box attacker has no special knowledge of the model other than being able to observe its predictions. On the other hand, a white-box attacker possesses full visibility into model architecture, parameters, ex., and importantly, can perform backpropagation to extract loss function gradients. The white-box scenario is generally viewed as the more insidious threat model. An untargeted attack tries to produce incorrect predictions on a malicious sample without regard to the exact nature of the miss-classification. Mathematically, given benign input $x$ with true label $y$, the untargeted attack optimizes for a perturbation $\delta$ that satisfies:

\begin{equation} \label{untargeted}
    \delta = \mathrm{argmax} \ \mathcal{L}(x+\delta ,y) \ \mathrm{s.t.} \ \|\delta\| < \epsilon
\end{equation}

\noindent where $\mathcal{L}$ is a loss function, and $\| \cdot \| $ is some norm. The norm constraint is needed to minimize the chance of detection. In contrast, a targeted attack seeks miss-classification as a \textit{specific} incorrect class. For benign input $x$ with associated label $y$, the adversary solves

\begin{equation} \label{targeted}
    \delta = \mathrm{argmin} \ \mathcal{L}(x+\delta ,\hat{y}) \ \mathrm{s.t.} \ \|\delta\| < \epsilon
\end{equation}

\noindent where $\hat{y}$ is the targeted label for the (incorrect) class the adversary wants the model to predict. Once again the constraint $\|\delta\| < \epsilon$ ensures that the attack is relatively imperceptible to the human user.

Some of the earliest work on evasion attacks was presented by \citet{biggio2013evasion}. They formulated the adversary's objective in terms of a constrained optimization problem in which the attacker searches for a perturbation that successfully `fools' the target model, while simultaneously ensuring that the attacked example remained within an $\epsilon$-bound of the original. A gradient descent approach was proposed to generate these adversarial examples. A similar framework was suggested by \citet{szegedy2013intriguing}, although L-BFGS was used in place of gradient descent. Additionally, \citet{szegedy2013intriguing} demonstrated the transferability of adversarial examples across deep learning models (i.e. attacks generated for one model can successfully fool a different one). \citet{goodfellow2014explaining} link attack transferability to excessive model linearity, hypothesizing that successful perturbations are highly aligned with weight vectors, and thus tend to generalize well. They also introduce the fast gradient sign method `FGSM' for quickly constructing untargeted adversarial examples. For an FGSM attack with an $\ell_{\infty}$ norm constraint of size $\epsilon$, the adversarial perturbation $\delta$ for sample $x$ is given by $\delta = \epsilon \ \mathrm{sign}(\nabla_x  J(x,y;\theta ))$. Here $J$ is some loss function, $y$ is the true label for $x$, and $\theta$ represents the model's parameters. \citet{madry2017towards} extended the FGSM methodology to introduce the Project Gradient Descent (PGD) attack. PGD employs an iterative approach to craft adversarial examples, with each step consisting of an FGSM perturbation followed by projection onto the $\epsilon$-ball. PGD attacks can be used in both targeted and untargeted threat models. 

\subsection{Evasion Attacks in Speech-centric ML}

A number of speech-specific attacks have also been proposed. \citet{gong2017crafting} provided one of the first investigations of end-to-end gradient-based white box attacks on audio-modality classifiers (e.g. speaker recognition). They demonstrated that the attacks substantially degrade model accuracy while only minimally impacting human perceptual evaluation. \citet{real_bob} also presented an attack against speaker recognition models but used a black-box approach that relies on gradient estimation. Black-box approaches to attacking ASR systems have been introduced by \citet{abdullah2021hear}, \citet{alzantot2018did}, and \citet{khare2018adversarial}. 

Meanwhile, \citet{carlini2018audio} presented a strong targeted white box attack against ASR, that obtains a perfect success rate (w.r.t. the ability of the attack to produce the targeted transcript) with greater than $30$ dB mean adversarial SNR. Their approach specified the generic adversarial framework for the Connectionist Temporal Classification (CTC) loss commonly used for training ASR systems. They also quantify perturbation magnitude in decibels instead of linear units. Within the framework from \citet{carlini2018audio}, attack generation equates to solving the following optimization problem 

$$\min |\delta|_2^2 + \alpha \mathcal{L}_\mathrm{CTC}(x+\delta, t) \ \mathrm{s.t.} \ \mathrm{dB}(\delta) - \mathrm{dB}(x) < \tau$$

\noindent where $t$ is the targeted transcription and $\mathcal{L}_\mathrm{CTC}$ is the CTC loss function. By optimizing this objective with successively smaller values of $\tau$, the adversary determines the smallest magnitude perturbation that achieves the targeted objective. This attack methodology was extended by \citet{qin2019imperceptible} which introduced the `Imperceptible attack' for ASR. After finding a perturbation that fools the network, an additional update step was used to regularize the power spectrum of the attack such that it falls under the masking threshold of the clean speech. These two steps (perturbation update and spectral shaping) were repeated for a number of iterations. In this manner, the Imperceptible attacks leveraged audio-specific notions of perceptibility in constraining the attack generation process. Similar work on psychoacoustically motivated attacks has also been presented in \citet{schonherr2018adversarial}. 

\subsection{Defenses against Evasion Attacks in Speech-centric ML}

A variety of defenses have been proposed to counteract the insidious effects of evasion attacks. One method that has been shown to be broadly successful is adversarial training (AT), in which adversarial examples are generated during the training process and used to further tune the model weights (\cite{goodfellow2014explaining, madry2017towards})\footnote{Note that \cite{madry2017towards} investigates both attacks (PGD) and defenses (AT)}. This approach is not unlike traditional data augmentation methods such as additive noise or image cropping, except that the adversarial samples are generated specifically for the model at hand (usually using a white-box attack such as FGSM). Unfortunately, despite its success, AT introduces a significant computational overhead since it requires the construction of adversarial examples on each training epoch, as well as additional weight updates.

Another widely known defense is Randomized Smoothing (RS), first introduced by \citet{cohen2019certified}. RS uses stochastic averaging to `wash-out' the effects of adversarial perturbations, which are relatively small in absolute magnitude. Specifically, given a classifier $f(\cdot)$ that is robust to additive Gaussian noise, a new `smoothed' classifier $g(\cdot)$ is produced by $g(x) =  \mathrm{argmax}_k \ P(f(x+\epsilon)=k)$ with $\epsilon \sim \mathcal{N}(0,\sigma^2 I)$. Random smoothing can provide provable robustness guarantees to $\ell_2$ bounded attacks (see \cite{cohen2019certified}).

Other defenses include changes to model architecture and training procedures. For example, a modified RELU activation function that constrains the maximum value of a given neuron was suggested by \cite{zantedeschi2017efficient}. Similarly, \citet{cisse2017parseval} introduced Parseval Regularization which was based on minimizing the Lipschitz constant. Other works, such as \citet{liao2018defense} and \citet{osadchy2017no}, have proposed the use of denoisers of other preprocessor methods to remove or mitigate the impact of adversarial perturbations.

Speech-specific defenses include the detection framework for ASR attacks from \citet{hussain2021waveguard} which leverages the sensitivity of adversarial examples to small changes. They identified adversarial samples by testing whether the predicted transcript varied substantially under alterations such as filtering and quantization. \citet{yang2018characterizing} used a similar approach for attack detection that tests the consistency between the transcript prediction for the full audio signal and the prediction for a partial segment. Several papers have also proposed audio resyntheses defenses, such as the GAN-based methods from \citet{esmaeilpour2021class} and \citet{esmaeilpour2021cyclic}.

\begin{table}[]
    \centering
    \caption{Summary of key works on adversarial attacks}
    \footnotesize
    \begin{tabular}{|p{1.5cm}|p{4cm}|p{4cm}|}

        \toprule
        \multicolumn{1}{c}{} & 
        \multicolumn{1}{c}{\textbf{Evasion Attacks}} & 
        \multicolumn{1}{c}{\textbf{Poisoning Attacks}} \\

        \midrule

        \multicolumn{1}{c}{} & 
        \multicolumn{1}{c}{{\citet{goodfellow2014explaining}}} & 
        \multicolumn{1}{c}{{\citet{biggio2012poisoning}}} \\

        \multicolumn{1}{c}{} & 
        \multicolumn{1}{c}{{\citet{madry2017towards}}} & 
        \multicolumn{1}{c}{{\citet{munoz2017towards}}} \\

        \multicolumn{1}{c}{\textbf{Attacks}} & 
        \multicolumn{1}{c}{{\citet{carlini2018audio}}} & 
        \multicolumn{1}{c}{{\citet{liu2017trojaning}}} \\

        \multicolumn{1}{c}{} & 
        \multicolumn{1}{c}{\textbf{}} & 
        \multicolumn{1}{c}{{\citet{gu2017badnets}}} \\

        \midrule

        \multicolumn{1}{c}{} & 
        \multicolumn{1}{c}{{\citet{goodfellow2014explaining}}} & 
        \multicolumn{1}{c}{{\citet{steinhardt2017certified}}} \\

        \multicolumn{1}{c}{\textbf{Defenses}} & 
        \multicolumn{1}{c}{{\citet{cohen2019certified}}} & 
        \multicolumn{1}{c}{{\citet{tran2018spectral}}} \\

        \multicolumn{1}{c}{} & 
        \multicolumn{1}{c}{{\citet{cisse2017parseval}}} & 
        \multicolumn{1}{c}{{\citet{gao2019strip}}} \\

        \bottomrule
        
    \end{tabular}
    
    \label{tab:adversarial attacks}
\end{table}

\section{Poisoning Attacks and Defenses}

\subsection{Preliminaries}

 Whereas an evasion attack occurs at inference time, poisoning attacks involve corruption of the training data thus leading to an inaccurate or vulnerable model. Given the increasing use of publicly sourced data and federated learning, it is often impossible to guarantee the integrity of a given dataset. The authors of \citet{barreno2010security} provided one of the earliest investigations of poisoning attacks. In particular, they distinguish between availability attacks that attempt to a broadly inaccurate model and integrity attacks that seek to produce a more specific vulnerability (e.g., misclassification in response to the presence of a specific trigger). In both scenarios, the attacker has the ability to manipulate the features and/or labels for some small fraction of the training data to decrease the performance of the learned model.

A variety of approaches have been proposed to generate availability-type poisoning attacks. In this case, the adversarial objective is to generate a set of poisoned samples that minimizes the learned model's performance on a withheld set of test samples. \citet{biggio2012poisoning} presented an attack on the SVM classifier that used gradient methods to generate the poisoned samples. SVM poisoning attacks have also been studied by \citet{mei2015using} in addition to attacks against linear and logistic regression. The Karush-Kuhn-Tucker (KKT) optimal conditions to generate the poisoned samples. While the relative simplicity of these models makes direct optimization feasible, Deep Neural Networks require different approaches. For example, the authors of \citet{munoz2017towards} introduced a method that estimates back-propagation through the entire training procedure to efficiently generated poisoned samples to maximize the validation loss in the final model. Another approach suggested by \citet{yang2017generative} was to generative poisoned samples with the Generative Adversarial Network (GAN) where the target model acted as the discriminator and a separate auto-encoder model was used as the generator that synthesized the poisoned samples.

A large number of integrity-type poisoning attacks attempt to produce a model that performs normally on clean test samples while miss-classifying those that contain a specific trigger. For example, \citet{liu2017trojaning} showed how a pre-trained facial recognition model can be `Trojaned' by creating a trigger-induced backdoor. They first used the model gradient to generate a trigger patch that produced a large-magnitude activation for some hidden layer neurons. Then they reconstructed faux training examples for each class by performing gradient ascent on the input image. These synthetic samples form the basis for a finetuning procedure, in which a subset of the samples have the trigger added and the label changed. The resulting model should then miss-classify future samples containing the trigger. Importantly, this is accomplished without direct knowledge of the training dataset. A similar threat model was introduced by \citet{gu2017badnets}, which demonstrated the efficacy of the approach by creating a poisoned model that incorrectly identified street signs when a small trigger was present.

While both \citet{liu2017trojaning} and \citet{gu2017badnets} assumed some attack autonomy over model training, other work such as by \citet{chen2017targeted}, assumed that the attacker only has access to a small percentage of the training data. The authors proposed several approaches to generate backdoor samples, including an input-instance-key method in which the poisoned samples are randomly perturbed versions of a `key' input with the target label. The adversary hoped that the final model would thus misclassify test samples that are similar to the key input used to poison the training data. Another poisoning algorithm suggested by \citet{liu2017trojaning} was to create poisoned instances by adding a specific pattern to benign samples, and changing the label to the targeted class. A clean-label attack that manipulated only the input features was presented by \citet{shafahi2018poison}. Their approach optimized for an adversarial sample that is perceptually similar to the target class in input space but close to a target instance in feature space. If the target instance is then deployed on the trained model, it is more likely to be misclassified as the target class. Another clean label attack was the Witches Brew method by \citet{geiping2020witches}, which used gradient matching to produce samples that modified the model's training trajectory such that it would misclassify a specific target instance at test time.

\subsection{Poisoning Attacks in Speech-centric ML}

So far, there has been relatively little work done on audio-specific poisoning attacks. The first targeted attack against hybrid ASR systems is the VenoMave algorithm by \citet{aghakhani2020venomave}. For each $x_i$ frame selected by the adversary for misclassification, that attacker constructs poisoned frames that `surround' $x_i$ in feature space and assigns them the target label. Thus any linear model that correctly classifies the poisoned frames will also misclassify the target frame. \citet{ge2022wavefuzz} proposed a clean-label attack to protect user's speech data from use in downstream learning tasks such as speaker recognition or speech command recognition. This was accomplished by introducing perturbations that maximize the distance between the MFCC features of the clean and poisoned signal while simultaneously minimizing the perturbation's magnitude. Thus any model trained on the poisoned data that uses MFCCs is likely to perform unreliably.

\subsection{Defenses against Poisoning Attacks in Speech-centric ML}
Many of the defenses proposed for poisoning attacks attempt to identify and remove the poisoned samples from the training data. For example, \citet{tran2018spectral} extracted learned representation for all training samples with a given label. They then compute a singular value decomposition (SVD) and use the right singular vectors to compute outlier scores for each sample. Samples with a high outlier score are removed, and the model is retrained. Model activations were also used to detect poisoned samples by \citet{chen2018detecting}, but the authors employ a clustering approach in place of the SVD. They also suggest fine-tuning the model on the re-labeled poisoned data rather than training from scratch on the filtered dataset. \citet{gao2019strip} leveraged the fact the sample contains a trigger that should be consistently classified as that target class even under perturbation. By randomly combining different images, the check for samples that exhibit low entropy in the distribution of predicted labels. Some defensive method focus on addressing backdoors in the model itself. For example, \citet{wang2019neural} tested for backdoors by computing the minimal perturbations required to change the classification result of all samples to a given target label. If a perturbation of a relatively small magnitude exists, then this may indicate the presence of a backdoor. The `Fine-pruning' defense presented by \citet{liu2018fine} first pruned neurons that were inactive on clean samples under the hypothesis that these extraneous neurons can be co-opted by the backdoor trigger. Then the network was fine-tuned on a clean (un-poisoned) dataset to avoid overall performance degradation. \citet{steinhardt2017certified} provided a method to upper bound the effects of a potential poisoning attack in the case where a data sanitation defense was used to remove outliers before training.

\chapter{Privacy in Speech-centric Machine Learning}

Despite the promises modern speech applications can deliver, they also raise significant concerns and risks, such as exposing sensitive information that people might wish to keep confidential. The sensitive information can be individual attributes (e.g., age, gender), states (e.g., health, emotions), or biometric fingerprints. This section presents a comprehensive review of the privacy and security challenges related to trustworthy speech processing.

\section{Taxonomies of Privacy-related Speech-centric ML}

\begin{figure*}[t]
    \begin{center}
    \tikzset{
      basic/.style  = {draw, text width=5cm, drop shadow, rectangle},
      root/.style   = {basic, rounded corners=3pt, thin, align=center, font=\huge\sffamily, fill=white, minimum height=2cm, text width=15cm, level distance=3cm},
      level-2/.style = {basic, rounded corners=3pt, thin, align=center, font=\LARGE\sffamily, fill=white, text width=3.5cm, fill=blue!20},
      level-3/.style = {basic, rounded corners=3pt, thin, align=center, font=\sffamily, fill=white, text width=2.4cm, fill=green!10},
      level-4/.style = {basic, thin, align=center, fill=white, font=\sffamily, text width=2cm, minimum height=0.75cm, fill=yellow!10}
    }

    \resizebox{\textwidth}{!}{
        \begin{tikzpicture}[
            basic/.style={draw, drop shadow, font=\sffamily, rectangle},
            level 1/.style={sibling distance=8.8cm, level distance=3.25cm, text width=3cm, minimum height=1.5cm},
            level 2/.style={sibling distance=2.8cm, level distance=3.25cm, minimum height=1.5cm},
            level 3/.style={sibling distance=2.8cm, level distance=3cm, minimum height=3cm},
            level 4/.style={sibling distance=1.0cm, level distance=0.1cm, minimum height=2cm},
            edge from parent/.style={->, solid,black,sloped,draw,line width=0.5mm}, 
            edge from parent path={(\tikzparentnode.south) -- (\tikzchildnode.north)},
            >=latex, node distance=2cm, edge from parent fork down, line width=0.3mm]
    
        \node[root] {\textbf{Privacy-enhancing\\Speech-centric Machine Learning}}
            child {
                node[level-2] (c1) {\textbf{Privacy Threats}}
                child {node[level-3] (c11) {\textbf{Property Inference Attack\\(PIA)}}}
                child {node[level-3] (c12) {\textbf{Membership Inference Attack\\(MIA)}}}
                child {node[level-3] (c13) {\textbf{Identity Inference Attack\\(IIA)}}}
                child {node[level-3] (c14) {\textbf{Content Inference Attack\\(CIA)}}}
            }
            child {
                node[level-2] (c2) {\textbf{Mitigation Mechanisms}}
                child {node[level-3] (c21) {\textbf{Algorithmic Solution}}}
                child {node[level-3] (c22) {\textbf{Hardware Solution}}}
            }
            child {
                node[level-2] (c3) {\textbf{Downstream Applications}}
                child {node[level-3] (c31) {\textbf{Automatic Speech Recognition\\(ASR))}}}
                child {node[level-3] (c32) {\textbf{Speech Emotion Recognition\\(SER)}}}
                child {node[level-3] (c33) {\textbf{Keyword Spotting}}}
                child {node[level-3] (c34) {\textbf{Speaker Verification}}}
            }
            child {
                node[level-2] (c4) {\textbf{Training Paradigms}}
                child {node[level-3] (c41) {\textbf{Centralized Learning}}}
                child {node[level-3] (c42) {\textbf{Federated Learning}}}
            };

            \begin{scope}[every node/.style={level-4}]
            \node [below of=c11, xshift=5pt] (c111) {\textbf{\cite{feng2022enhancing}}};
            \node [below of=c111] (c112) {\textbf{\cite{feng2021privacy}}};
            \node [below of=c112] (c113) {\textbf{\cite{jaiswal2020privacy}}};
            \node [below of=c113] (c114) {\textbf{\cite{stoidis2022generating}}};
            \node [below of=c114] (c115) {\textbf{\cite{muller22b_interspeech}}};
            \node [below of=c115] (c116) {\textbf{\cite{aloufi2020privacy}}};
            \node [below of=c116] (c117) {\textbf{\cite{aloufi2019emotionless}}};

            \node [below of=c12] (c121) {\textbf{\cite{tseng22_interspeech}}};
            \node [below of=c121] (c122) {\textbf{\cite{miao2019audio}}};
            \node [below of=c122] (c123) {\textbf{\cite{shah21_interspeech}}};
            
            \node [below of = c13] (c131) {\textbf{\cite{han2020voice}}};
            \node [below of = c131] (c132) {\textbf{\cite{pathak2012privacy}}};
            \node [below of = c132] (c133) {\textbf{\cite{tomashenko2022voiceprivacy}}};
            \node [below of = c133] (c134) {\textbf{\cite{pierre22_interspeech}}};
            \node [below of = c134] (c135) {\textbf{\cite{srivastava20_interspeech}}};
            \node [below of = c135] (c136) {\textbf{\cite{granqvist20_interspeech}}};
            \node [below of = c136] (c137) {\textbf{\cite{maouche20_interspeech}}};
            \node [below of = c137] (c138) {\textbf{\cite{nelus19_interspeech}}};
            \node [below of=c138] (c139) {\textbf{\cite{qian2018towards}}};
            \node [below of=c139] (c1310) {\textbf{\cite{stoidis2022generating}}};

            \node [below of = c14] (c141) {\textbf{\cite{ahmed2020preech}}};
            \node [below of = c141] (c142) {\textbf{\cite{amid22_interspeech}}};
            \node [below of = c142] (c143) {\textbf{\cite{liu22_interspeech}}};

            \node [below of = c21] (c211) {\textbf{\cite{huang22k_interspeech}}};
            \node [below of = c211] (c212) {\textbf{\cite{jaiswal2020privacy}}};
            \node [below of = c212] (c213) {\textbf{\cite{zheng2022keyword}}};
            \node [below of = c213] (c214) {\textbf{\cite{thebaud2021spoofing}}};
            \node [below of = c214] (c215) {\textbf{\cite{leroy2019federated}}};
            \node [below of = c215] (c216) {\textbf{\cite{hard22_interspeech}}};
            \node [below of = c216] (c217) {\textbf{\cite{pathak2012privacy}}};
            \node [below of = c217] (c218) {\textbf{\cite{tomashenko2020voiceprivacy}}};
            \node [below of = c218] (c219) {\textbf{\cite{tomashenko2022voiceprivacy}}};
            \node [below of = c219] (c2110) {\textbf{Rahulama\\thavan \textit{et al.} 2018}};

            \node [below of = c22] (c221) {\textbf{\cite{brasser18_interspeech}}};
            \node [below of = c221] (c222) {\textbf{\cite{kumar2015sound}}};
            \node [below of = c222] (c223) {\textbf{\cite{bayerl2020offline}}};

            \node [below of = c31] (c311) {\textbf{\cite{amid22_interspeech}}};
            \node [below of = c311] (c312) {\textbf{\cite{pierre22_interspeech}}};
            \node [below of = c312] (c313) {\textbf{\cite{huang22k_interspeech}}};
            \node [below of = c313] (c314) {\textbf{\cite{chennupati2022ilasr}}};
            \node [below of = c314] (c315) {\textbf{\cite{tomashenko2022privacy}}};
            \node [below of = c315] (c316) {\textbf{\cite{zhu22b_interspeech}}};
            \node [below of = c316] (c317) {\textbf{\cite{amid22_interspeech}}};
            \node [below of = c317] (c318) {\textbf{\cite{aloufi21_interspeech}}};
            \node [below of = c318] (c319) {\textbf{\cite{srivastava19_interspeech}}};
            \node [below of = c319] (c3110) {\textbf{\cite{brasser18_interspeech}}};
    
            \node [below of = c32] (c321) {\textbf{\cite{feng2022enhancing}}};
            \node [below of = c321] (c322) {\textbf{\cite{feng2021privacy}}};
            \node [below of = c322] (c323) {\textbf{\cite{jaiswal2020privacy}}};
            \node [below of = c323] (c324) {\textbf{\cite{arora2018exploring}}};
            \node [below of = c324] (c325) {\textbf{\cite{tsouvalas2022privacy}}};
            \node [below of = c325] (c326) {\textbf{\cite{feng2022semi}}};
            \node [below of = c326] (c327) {\textbf{\cite{dias2018exploring}}};

            \node [below of=c33] (c331) {\textbf{\cite{zheng2022keyword}}};
            \node [below of=c331] (c332) {\textbf{\cite{elworth2022hekws}}};
            \node [below of=c332] (c333) {\textbf{\cite{liu2020datamix}}};
            \node [below of=c333] (c334) {\textbf{\cite{leroy2019federated}}};
            \node [below of=c334] (c335) {\textbf{\cite{hard2020training}}};
            \node [below of=c335] (c336) {\textbf{\cite{hard22_interspeech}}};

            \node [below of = c34] (c341) {\textbf{\cite{noe21_interspeech}}};
            \node [below of = c341] (c342) {\textbf{\cite{pathak2012privacy}}};
            \node [below of = c342] (c343) {\textbf{\cite{granqvist20_interspeech}}};
            \node [below of = c343] (c344) {\textbf{Rahulama\\thavan \textit{et al.} 2018}};
            \node [below of = c344] (c345) {\textbf{\cite{treiber2019privacy}}};
            \node [below of = c345] (c346) {\textbf{\cite{smaragdis2007framework}}};
            \node [below of = c346] (c347) {\textbf{\cite{chouchane21_interspeech}}};

            \node [below of = c41] (c411) {\textbf{\cite{amid22_interspeech}}};
            \node [below of = c411] (c412) {\textbf{\cite{feng2022enhancing}}};
            \node [below of = c412] (c413) {\textbf{\cite{jaiswal2020privacy}}};
            \node [below of = c413] (c414) {\textbf{\cite{zhu22b_interspeech}}};
            \node [below of = c414] (c415) {\textbf{\cite{huang22k_interspeech}}};
            \node [below of = c415] (c416) {\textbf{\cite{tseng22_interspeech}}};
            \node [below of = c416] (c417) {\textbf{\cite{stoidis2022generating}}};
            \node [below of = c417] (c418) {\textbf{\cite{aloufi21_interspeech}}};
            \node [below of = c418] (c419) {\textbf{\cite{han2020voice}}};
    
            \node [below of = c42] (c421) {\textbf{\cite{tsouvalas2022privacy}}};
            \node [below of = c421] (c422) {\textbf{\cite{feng22b_interspeech}}};
            \node [below of = c422] (c423) {\textbf{\cite{zhang2022fedaudio}}};
            \node [below of = c423] (c424) {\textbf{\cite{leroy2019federated}}};
            \node [below of = c424] (c425) {\textbf{\cite{yu2021federated}}};
            \node [below of = c425] (c426) {\textbf{\cite{granqvist20_interspeech}}};
            \node [below of = c426] (c427) {\textbf{\cite{zhu22b_interspeech}}};
            \node [below of = c427] (c428) {\textbf{\cite{cui2022privacy}}};
            \node [below of = c428] (c429) {\textbf{\cite{hard22_interspeech}}};

            \end{scope}
        
            \foreach \value in {1,2,3,4,5,6,7}
              \draw[->] (c11.195) |- (c11\value.west);
    
            \foreach \value in {1,2,3}
              \draw[->] (c12.195) |- (c12\value.west);
    
            \foreach \value in {1,2,3,4,5,6,7,8,9,10}
              \draw[->] (c13.195) |- (c13\value.west);
    
            \foreach \value in {1,2,3}
              \draw[->] (c14.195) |- (c14\value.west);
    
            \foreach \value in {1,2,3,4,5,6,7,8,9,10}
              \draw[->] (c21.195) |- (c21\value.west);
    
            \foreach \value in {1,2,3}
              \draw[->] (c22.195) |- (c22\value.west);
    
            \foreach \value in {1,2,3,4,5,6,7,8,9,10}
              \draw[->] (c31.195) |- (c31\value.west);
              
            \foreach \value in {1,2,3,4,5,6,7}
              \draw[->] (c32.195) |- (c32\value.west);
    
            \foreach \value in {1,2,3,4,5,6}
              \draw[->] (c33.195) |- (c33\value.west);
    
            \foreach \value in {1,2,3,4,5,6,7}
              \draw[->] (c34.195) |- (c34\value.west);
    
            \foreach \value in {1,2,3,4,5,6,7,8,9}
              \draw[->] (c41.195) |- (c41\value.west);
              
            \foreach \value in {1,2,3,4,5,6,7,8,9}
              \draw[->] (c42.195) |- (c42\value.west);
        \end{tikzpicture}
    }
    
    \end{center}
    \caption{Taxonomy of privacy-related speech processing and modeling works.}
    \label{fig:privacy-taxonomy}
\end{figure*}
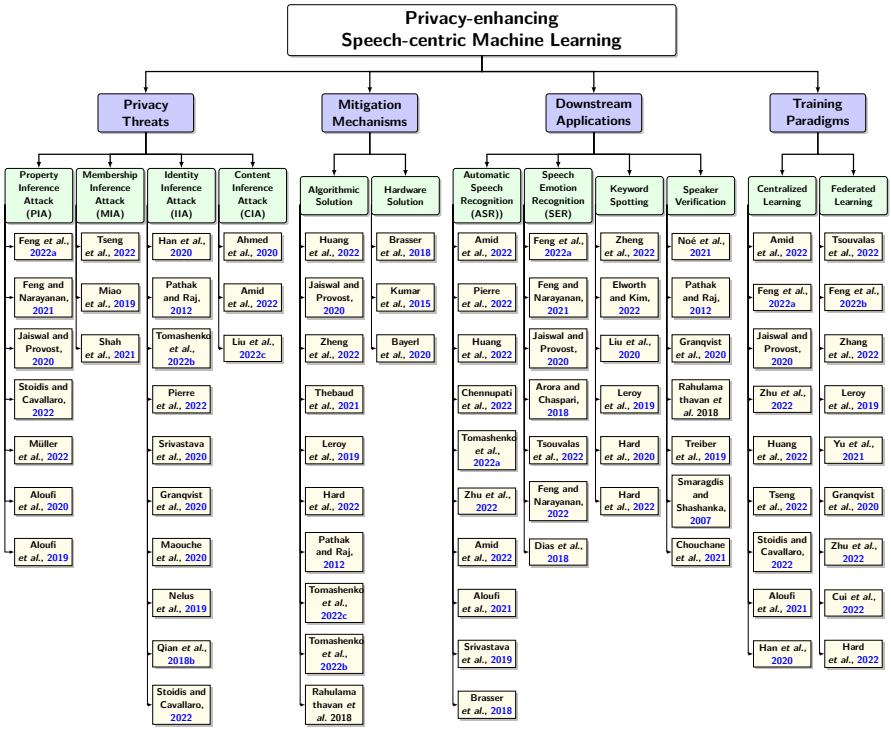

In this subsection, we create a taxonomy of existing privacy-related topics on speech-centric ML in Figure~\ref{fig:privacy-taxonomy}. In the categorization shown in Figure~\ref{fig:privacy-taxonomy}, we organize papers based on privacy threats, mitigation mechanisms, downstream applications, and training paradigms. In the first place, we categorize privacy threats based on privacy attacks, including Property Inference Attacks (PIA), Membership Inference Attacks (MIA), Identity Inference Attacks (IIA), and Content Inference Attacks (CIA). For example, in PIA, the privacy attacker can obtain or infer speaker attributes like demographic information. On the other hand, we can divide privacy-related literature based on privacy mitigation methods. Specifically, the literature surrounding privacy-preserving speech processing can be categorized into algorithmic solutions and hardware solutions. Moreover, the most studied downstream speech applications related to privacy topics can be categorized into automatic speaker verification, keyword spotting, automatic speech recognition, and speech emotion recognition. Lastly, we discuss the privacy-related speech-centric ML based on the training paradigm: centralized learning and federated learning (FL).

\begin{figure*}[t]
    \begin{center}
        \includegraphics[width=\linewidth]{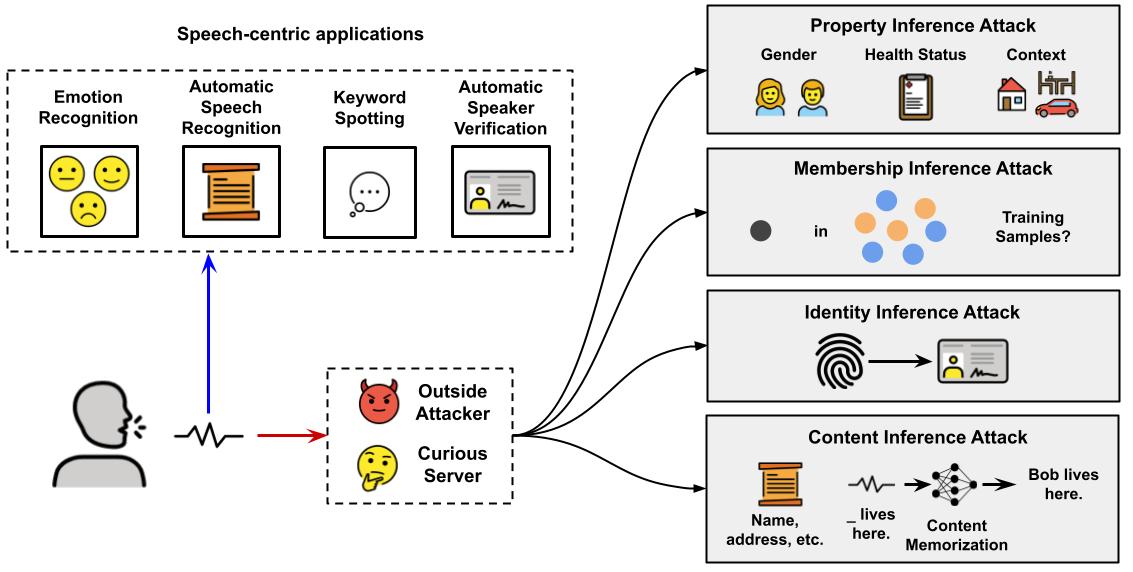}
    \end{center}
    \caption{Overview of the privacy threats in speech-centric models.}
    \label{fig:privacy_threats}
\end{figure*}

\section{Privacy Threats}

In a privacy attack, the goal of an adversary is to acquire information that was not intended to be disclosed, including speech content, speaker demographics, and voice fingerprint. There are four mainstream privacy attacks against speech-centric applications as demonstrated in Figure~\ref{fig:privacy_threats}: Property Inference Attacks (PIA), Membership Inference Attacks (MIA), Identity Inference Attacks (IIA), and Content Inference Attacks (CIA). We provide a brief overview of each privacy attack and highlight works that either identified new privacy risks or were the first to investigate privacy threats in their respective domains.

\begin{figure*}[t]
    \begin{center}
        \includegraphics[width=\linewidth]{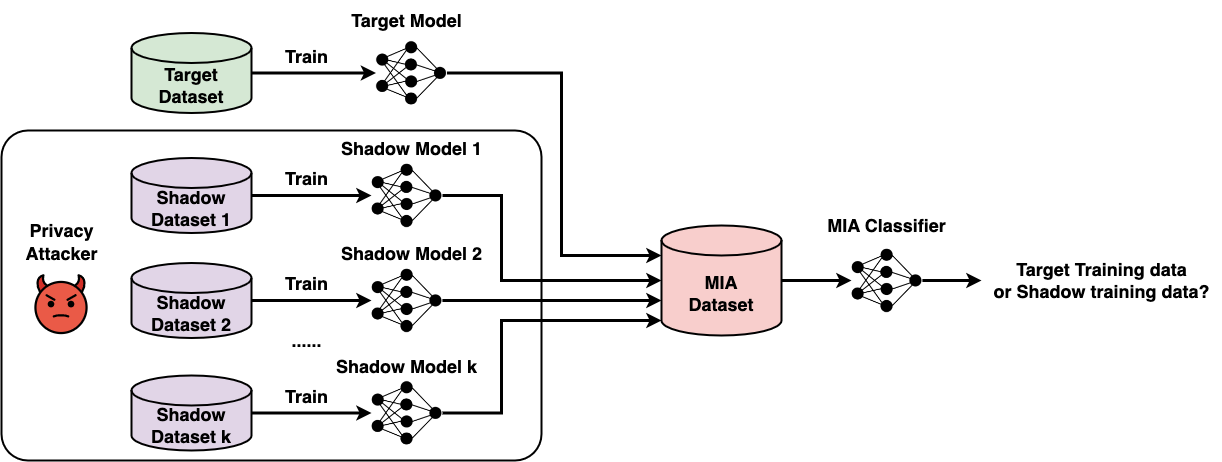}
    \end{center}
    \caption{Overview of the Membership inference attack. The privacy attacker first trains a set of shadow models with shadow training datasets. Once the shadow training is finished, the attacker gathers the model outputs from the target model and the shadow model to train the MIA classifier. The MIA classifier infers the membership property given an input posterior from the target model.}
    \label{fig:mia}
\end{figure*}

\vspace{1mm}
\noindent \textbf{Property Inference Attacks (PIA)}: PIA occurs when the adversary attempts to infer private attributes which are unrelated to the primary learning task. A notable example of the PIA in speech-centric applications is to infer the gender attribute using a pre-trained gender classification model, while the target application is to classify emotions or transcribe text. Here, the speech data that is accessible to the attacker can either be the raw speech recordings or processed speech features like MFCCs. In addition to gender property, adversaries can perform classification to predict the age (\cite{safavi2018automatic}), the language used (\cite{lopez2014automatic}), or even the health status (\cite{akman2022evaluating}) of the speaker from the speech data. However, since most existing speech-related datasets only include the annotations of gender but not other properties, the majority of the PIA works in speech applications focus on gender classification. Furthermore, apart from using features derived from raw speech data, the adversaries could perform privacy attacks through training updates generated in the collaborative training process presented by \citet{feng22b_interspeech}.

\vspace{1mm}
\noindent \textbf{Membership Inference Attacks (MIA)}: Membership inference aims to determine the participation of a data instance in training the target model. The idea of the MIA was first proposed by \citet{shokri2017membership}, where the author assumed the attacker could access posterior estimations of a data sample by querying the target model. The attacker then used the posterior distribution to speculate whether the query data was in the training data. The general framework of MIAs is presented in Figure~\ref{fig:mia}. In MIAs, the attacker typically starts with a training procedure called shadow training which emulates the target training procedure. To perform the shadow training, the attacker often gathers a collection of shadow training datasets that share similar data distribution or data format to the target training data. The attacker then trains a classifier to perform MIAs using a collection of posteriors from training and shadow data. Although MIAs have been widely studied in computer vision and natural language processing, there is a limited amount of work in speech modeling. In speech-centric ML, \citet{shah21_interspeech} were the first to investigate MIAs on ASR models and their results show that the attacker can infer membership of the speech data with a moderate precision score. Furthermore, it is important to point out that the success rate of MIAs depends largely on the speakers, while some speakers are more vulnerable to MIAs. However, the connection between the attack success rate of MIAs and the speaker remains to be determined. Recently, \citet{tseng22_interspeech} designed the MIA against pre-trained speech models trained using self-supervised learning (SSL) in black-box settings. Their results indicate that both utterance and speaker-level MIAs are feasible against SSL-based speech models.

\vspace{0.5mm}
\noindent \textbf{Identity Inference Attacks (IIA)}: The IIA against speech-centric applications is a class of privacy attacks where adversaries can extract personally identifiable information (PII) from speech data for re-identification or impersonation purposes. However, we highlight that such identifiers often have applications in speaker identification and automatic speaker verification tasks. Traditional speaker identification frameworks rely on the extraction of i-vector from MFCCs (\cite{dehak2010front}). i-vector is a low-dimensional fixed-length representation of a speech utterance extracted using a data-driven approach. The state-of-art speaker identification systems in more recent years involve the extraction of the x-vector (\cite{snyder2018x}), which is the embedding extracted from Deep Neural Networks. Many papers (\cite{han2020voice, pathak2012privacy, tomashenko2022privacy, pierre22_interspeech, srivastava19_interspeech, nelus19_interspeech, maouche20_interspeech}) have proposed various privacy mitigation methods to prevent speaker re-identification. The vast majority of these works thus far applied adversarial training to disentangle the unique speaker information embedded in the speech recordings.

\vspace{0.75mm}
\noindent \textbf{Content Inference Attacks (CIA)}: In addition to speaker properties, speaker memberships, and biometric identifiers, the textual contents of the speech can expose significant privacy concerns. For example, speech recordings often contain the name, address, and contact information of the speaker or the people around the speaker. Moreover, speech recorded in sensitive settings like business meetings can include proprietary information. Therefore, the content inference is to extract textual content from speech recordings. The naive approach to performing the content inference is through ASR itself. More recently, researchers have discovered that untended memorization in training deep speech models can also leak training data records (\cite{amid22_interspeech, liu22_interspeech}). Specifically, the attacker can deliberately synthesize speech contents by guessing content patterns in the training data. For example, the crafted speech utterance can be "\_ lives in the 1st street." where \_ can be the silence audio snippet. The idea of the attack is that the ASR model would output the name along with the phrase "lives in the 1st street.", like "Bob lives in the 1st street.", as a consequence of model memorization.

\section{Mitigation Mechanisms}

In this paragraph, we continue our review of privacy-related literature on speech-centric ML based on mitigation strategies. We split the related papers into algorithmic and hardware solutions based on such criteria. 

\vspace{1mm}
\subsection{Algorithmic Solutions}

There is a large body of work falling into algorithmic solutions. More concretely, differential privacy (DP), adversarial training, and encryption are frequently adopted mitigation methods in speech-centric applications.

\subsubsection{Differential Privacy (DP)}
The idea of DP was first introduced by \citet{dwork2008differential}. Essentially, the DP is a rigorous privacy definition that guarantees the exclusion or the inclusion of any particular data record in the dataset has a negligible impact on the original data distribution. In other words, the privacy attacker cannot distinguish the data distribution changes by including or excluding any particular data point in the original dataset. Mathematically speaking, we can define DP given privacy parameters $\epsilon$ and $\delta$ shown below:

\begin{definition}[$(\epsilon, \delta)$-DP]
    A random mechanism $\mathcal{M}$ satisfies $(\epsilon, \delta)$-LDP, where $\epsilon>0$ and $\delta\in[0, 1)$, if and only if for any two adjacent data sets $\mathcal{D}$ and $\mathcal{D'}$ in universe $\mathcal{X}$, we have:
    
    \vspace{-1.5mm}
    \begin{equation}
        Pr(\mathcal{M}(\mathcal{D})) \leq e^{\epsilon} Pr(\mathcal{M}(\mathcal{D'})) + \delta
    \end{equation}
\end{definition}

Here, $\epsilon>0$ is defined as the privacy budget in DP, and a lower $\epsilon$ represents stronger privacy protection (\cite{dwork2006calibrating}). Specifically, $\epsilon$ provides the bound of all outputs on neighboring data sets $\mathcal{D}$ and $\mathcal{D'}$, which differ by one sample in a database. The typical way to achieve differential privacy is through noise perturbation. When considering centralized speech applications, \citet{qian2018hidebehind} proposed VoiceMask that conceals the voiceprints of the speaker using differential privacy. Unlike the DP implementation in VoiceMask, Preech (\cite{ahmed2020preech}) ensured DP by adding dummy words in the output transcripts. Besides the literature on centralized training, we have observed an increasing trend of DP research in Federated Learning. For example, \citet{feng22b_interspeech} applied the DP to mitigate property inference attackers in training the FL-based speech emotion recognition model. Sotto Voce (\cite{shoemate2022sotto}) was another recently proposed work that explored DP in Federated speech recognition.

\subsubsection{Adversarial Training} 
As \citet{mireshghallah2020privacy} pointed out, adversarial training is based upon information-theoretic privacy. The fundamental theory behind adversarial training is mutual information. For example, given a sensitive property or unique identifier $z$ and the associated speech data $\mathbf{x}$, we want to learn the perturbation $h(\cdot)$ that maximizes the mutual information ${I(h(\mathbf{x}); z)}$. This learning objective is typically turned into the following adversarial training objectives as suggested by \citet{song2019learning}:

\vspace{-1.5mm}
\begin{equation}
    \min_{\mathbf{\psi}} \max_{\mathbf{\phi}} \; \mathcal{L}(adv_{\mathbf{\psi}}(h_{\phi}(\mathbf{x})), z)
    \label{equ:adv_loss}
\end{equation}

Essentially, we would like to train an adversary that is able to infer $z$ accurately, and meanwhile, we aim to improve the quality of the perturbation that confuses the adversary classifier. This training objective is normally combined with the target training objective in the learning phase. As we described in PIAs, many papers used adversarial training to disentangle the gender attribute from the speech signal. \citet{jaiswal2020privacy} were the first ones to propose to use of adversarial training to remove gender property in the SER task. Following \citet{jaiswal2020privacy}, \citet{feng2022enhancing} combined adversarial training with feature selection to greatly reduce the gender inference risks in SER. Another common approach to conducting adversarial training is through generating adversarial examples. For instance, in ASR training, \citet{stoidis2022generating} presented a generative adversarial network that fed gender-ambiguous training samples to train the ASR model. This design attempted to disentangle gender from training utterances. Aside from unlearning demographics from speech signals, \citet{chennupati2022ilasr} tried to unlearn PII such as x-vector by synthesizing speech utterances from a large pool of x-vector.

\subsubsection{Encryption}
The last popular privacy-defending mechanism in this category is encryption. Many papers exploit Homomorphic Encryption (HE) for secure training of speech-centric models. \citet{elworth2022hekws} and \citet{zheng2022keyword} investigated the use of HE in keyword spotting systems, and \citet{dias2018exploring} were the first to investigate HE in SER applications. On top of HE integrations, \citet{pathak2012privacy} adapted secure multiparty computation (SMC) protocols that substantially reduce the computation overhead needed to satisfy the privacy constraints. However, in our literature search, we cannot find related papers investigating HE in the ASR system. The lack of ASR research in this direction can be caused by the heavy computation required in encryption computation. Meanwhile, the modern ASR system demands a significant amount of computing resources.

\subsection{Hardware Solutions}

Compared with algorithmic mitigation in speech-centric applications, fewer research papers work on hardware solutions. Based on the sampled literature, we divide the hardware solution into sensing strategies and trust computing environments. In the context of audio sensing, \citet{kumar2015sound} aimed to improve the privacy of audio data recorded from wearables by audio subsample and audio shredding on the device. Specifically, audio shredding was to randomize the sequence of recorded audio features. These two sensing strategies promised to provide useful audio features that were secure from reconstruction attacks. Additionally, \citet{feng2018tiles} introduced a wearable audio solution that enhances privacy through sampling low-level acoustic characteristics instead of raw audio samples to study workplace stress (\citet{mundnich2020tiles, yau2022tiles}). On the other hand, there has been a growing interest in recent years in using a trusted computing environment (TEE) in speech-centric applications. For example, VoiceGuard by \citet{brasser18_interspeech} was one of the first works demonstrating the use of Intel SGX, a widely available TEE implementation, on the ASR application. Last but not least, \citet{bayerl2020offline} built a TEE architecture called Offline Model Guard (OMG) that allowed running KWS tasks on the pre-dominant mobile computing platform ARM.

\section{Downstream Speech Applications} 

Here, we review the privacy-related speech literature based on the downstream applications. In this review, we select popular speech applications, including automatic speech recognition (ASR), speech emotion recognition (SER), automatic speaker verification (ASV), and keyword spotting (KWS). 

\vspace{1mm}
\noindent \textbf{ASR}: As shown in Figure~\ref{fig:privacy-taxonomy}, ASR is the most studied speech application in the context of privacy. These works implement privacy-enhancing features by removing gender property (\cite{aloufi2020privacy}), biometric identifiers (\cite{pierre22_interspeech}), and sensitive content (\cite{amid22_interspeech}) from speech signals. Specifically, the scientific community has also introduced VoicePrivacy 2020 (\cite{tomashenko2020voiceprivacy}) and VoicePrivacy 2022 challenges (\cite{tomashenko2022voiceprivacy}), with the target to evaluate privacy-preserving ASR modeling frameworks that suppress biometric identifiers in the speech signal. Currently, most of the privacy-enhancing ASR systems are proposed in the centralized setting, and Federated ASR training remains a challenge in speech-centric ML research. In this survey, we find only a few presented works (\cite{zhu22b_interspeech, yu2021federated}) that focus on ASR modeling using federated learning. As \citet{yu2021federated} concludes, ASR models suffer significant utility loss using Federated learning due to the nature of the complexity and high variability residing in speech data. The lack of ASR modeling works in the FL domain can also be caused by the expensive computation requirements of the ASR models. Unlike the algorithmic solutions to reduce privacy risks, \citet{brasser18_interspeech} introduced the VoiceGuard architecture that protects user privacy using inside a trusted execution environment (TEE). Last, we also want to highlight that the Librispeech (\cite{panayotov2015librispeech}) dataset is commonly used in privacy-related ASR works.

\vspace{1mm}
\noindent \textbf{SER}: In our review, we identify that many papers (\cite{jaiswal2020privacy, feng2021privacy, feng2022enhancing}) in this domain focus on disentangling the gender attribute from the speech signal using adversarial training. Among all these papers, \citet{jaiswal2020privacy} was the only work that considers multi-modal learning with other modalities. As opposed to gender obfuscation works mentioned above, \citet{arora2018exploring} was the first to explore speaker anonymization using siamese neural network architecture. Apart from centralized speech emotion recognition, many recent works (\cite{feng2022semi, tsouvalas2022privacy, feng22b_interspeech}) explored privacy risks in federated learning settings. Specifically, the IEMOCAP dataset (\cite{busso2008iemocap}) is one of the most used datasets in conducting these experiments.

\vspace{1mm}
\noindent \textbf{KWS}: The literature of privacy-preserving keyword spotting systems (\cite{zheng2022keyword, elworth2022hekws}) are mainly based on homomorphic encryption (HE) solutions. However, as we discussed earlier, this method has major constraints in computation efficiency and is extremely challenging to deploy in the field. Alternatively, DataMix by \citet{liu2020datamix} improves privacy by generating data samples through the mixup approach (\cite{zhang2017mixup}). The mixup data is a mixture of data samples that can effectively prevent privacy attacks like IIA while preserving the target model utility.

\vspace{1mm}
\noindent \textbf{ASV}: Interestingly, most privacy-centered speech modeling papers treat speaker identity as sensitive information. Most of these papers heavily studied the obfuscation of the speaker identity, or in other words, reduce the performance of the speaker verification system, where the target application is often ASR, SER, or other speech-related applications. Since the i-vector or the x-vector already carries the biometric identifier of the speaker, many works that attempt to preserve privacy in the speaker verification systems focus on redesigning the system using homomorphic encryption solutions (\cite{smaragdis2007framework, pathak2012privacy}). However, the homomorphic encryption frameworks require heavy computations and are frequently impractical for real-life deployment. In contrast to these holomorphic proposals, \citet{rahulamathavan2018privacy} designed a randomization algorithm that significantly reduced the computation overhead for privacy-enhancing speaker verification using the i-vector. With the popularity of Federated Learning in more recent years, \citet{granqvist20_interspeech} investigated the on-device learning schemes for local speaker verification.

\section{Training Paradigms}

\subsection{Centralized} 
\label{sec:privacy_central}
Centralized learning requires collecting the raw speech data. In this setting, the speech signal is normally sampled at the client device and is then transferred to the service provider's server for post-processing. The collection of speech data often draws substantial privacy concerns as speech signals encapsulate demographics, health information, PII, or sensitive speech content. If service providers are untrusted, they may not only infer the mentioned private information from speech data but also even render a person identifiable information. 

To decrease the privacy risk of gathering speech signals, many privacy regulations have been issued in recent years, such as \textit{European Union’s General Data Protection Regulation (GDPR)} law (\cite{voigt2017eu}). Noticeably, \textit{GDPR} does not explicitly consider the machine learning models as personal data, but recent works imply that ML models themselves could be covered by \textit{GDPR} as models may memorize sensitive information during the training process. With regard to the research community, a large amount of effort has been made to decrease the privacy risks in centralized speech systems like ASR, SER, and ASV. Many of these papers, as summarized earlier, focus on perturbing the speech data or intermediate speech features to disentangle private information. Widely used speech data perturbation methods are differential privacy, adversarial training, or generative methods. We would also stress that most current works emphasize hiding/removing PII, but fewer works studied the topic of preventing the rendering of PII.

\subsection{Federated Learning}
\begin{figure*}[t]
    \begin{center}
        \includegraphics[width=\linewidth]{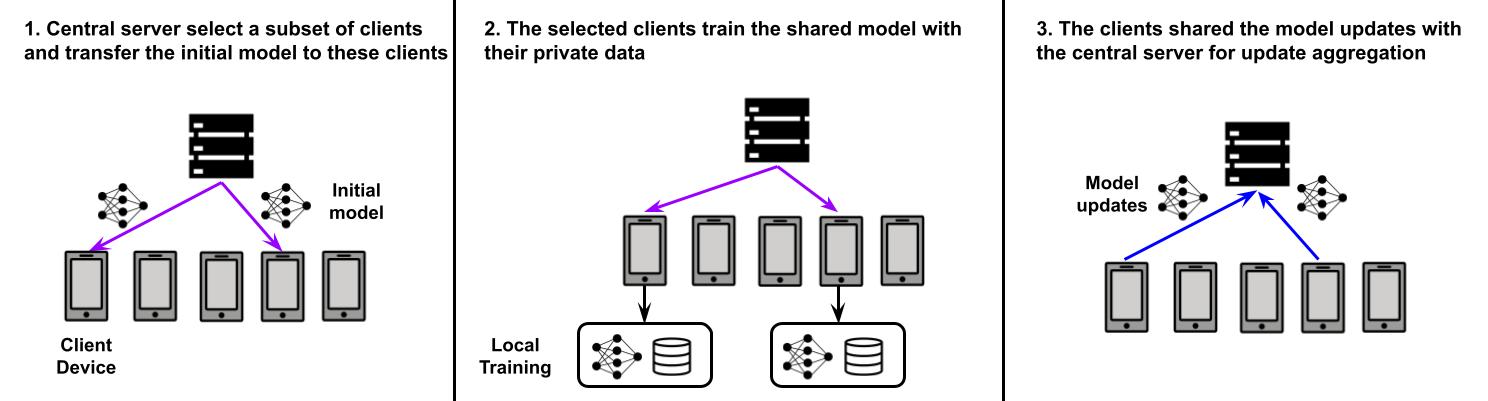}
    \end{center}
    \caption{The training process of Federated Learning.}
    \label{fig:fed_learning}
\end{figure*}

Section \ref{sec:privacy_central} has introduced the plethora of privacy risks centralized computing poses among which a significant attack concept is PII being transferred to a centralized server, wherein it is prone to cybersecurity attacks or curious server attacks. As an alternative to the traditional methods of training machine learning on a single server, federated learning (FL) employs a server-client model such that the data never leaves the client. Instead, the objective of the server is to aggregate all the model updates from the clients. Unlike centralized training, the clients train the model on a local dataset and transmit the updated parameters instead of the raw data. The general learning procedure of FL was shown in Figure~\ref{fig:fed_learning}.

\noindent\textbf{FL Optimization techniques} The earliest optimization algorithm proposed to ensure convergence of the global model was done by \citet{mcmahan2017communication}, known as Federated Averaging (\textit{FedAvg}). \textit{FedAvg} is similar to SGD, wherein the client model performs multiple iterations of updates before communicating the updates to the server. Although \textit{FedAvg} has been shown to have great success, it has convergence issues in heterogeneous data settings, i.e., when the data distribution is non-IID and in cases wherein a limited number of clients participate in every global update. One method proposed to improve optimization for heterogeneous data is the Stochastic Controlled Averaging Algorithm by \citet{karimireddy2020scaffold} (SCAFFOLD) by using control variates which reduce the client drift away from the global optima. More recently, \citet{asad2020fedopt} proposed three adaptive optimization algorithms, FedAdaGard, FedAdam, and FedYOGI, which are the FL versions of AdaGard, Adam, and YOGI, respectively. From empirical estimates, FedOpt outperforms most federated optimization strategies. Federated learning for speech processing tasks has been explored for numerous applications, mainly in keyword spotting, automatic speech recognition, and speech emotion recognition. Specifically, \citet{zhang2022fedaudio} provided a comprehensive benchmark for various audio-related tasks.

\noindent\textbf{KWS} With the ubiquitous usage of smart assistants such as Siri, Google Voice, and Alexa, keyword spotting is an essential downstream speech processing task. Furthermore, it requires a relatively low parameter count ($\sim$ 200k) to achieve state-of-the-art performance, making it ideal for federated learning. \citet{leroy2019federated, hard2020training} proposed the FL approach for keyword spotting tasks. Both papers provided models with comparable performance to a centrally trained model on their respective datasets. \citet{hard2020training} observed that the performance of the FL model depends on the strategies employed to deal with the non-IID data. For example, data augmentation through SpecAug (\cite{park2019specaugment}) or the usage of adaptive optimization methods such as ADAM in the local clients have been observed to be effective in preserving performance. In more recent work, \citet{hard22_interspeech} demonstrated a real-time on-device training of an FL model for Keyword Spotting. Also, it introduces semi-supervised learning in an FL scenario and self-correcting labels based on metadata during the recordings.

\noindent\textbf{ASR}
\citet{dimitriadis2020federated} was one of the earliest works to propose FL for Speech Recognition tasks. In order to train with heterogeneous data, a two-level hierarchical optimization strategy was proposed, which involved a local client optimization followed by a global optimization and retraining of the global model on the client side held out dataset. This helps the training process better adjust to client drift. In addition, a weight model averaging is proposed, which helps improve the convergence speed. In order to account for data heterogeneity, the addition of variational noise has been proposed by \citet{guliani2021training} wherein each client model is added with a local random variational vector. The author named this method as federated variational noise (FVN). FVN has been shown to improve the relative performance of the federated learning models in a non-IID data scenario.

With the constraints on the computational costs and the model sizes, there has been a focus on employing a  cross-silo FL framework. Since the number of clients is smaller and each client has much larger computing power than a cross-device setting, usage of large-scale ASR models is justifiable. \citet{cui2021federated} proposed a cross-silo FL framework that contained a detailed analysis of training an ASR system with multi-domain data, i.e., each client has a specific domain exclusive data such as read speech, conversational data, meeting data etc. It introduces the client-adaptive federated learning (CAFT) method which accounts for the differing domain modalities across clients and adapts the client's data using a transform. \citet{nandury2021cross} experimented a modification to FedAvg termed as FedAvg-DS wherein DS stands for Diversity scaling. This modification emphasizes accounting for the variability in gradient directions of the local client updates.

Recent efforts have enabled cross-device FL-based ASR training by \citet{guliani2022enabling}, which employed federated dropout (\cite{caldas2018expanding}) in order to reduce the model sizes in the clients and at the same time obtain a fully trained ASR at the server side. Additionally, it has been shown that training with federated dropout allows sub-models of the fully trained model to have comparable performance allowing for deployment on devices with varying computing capacities. \citet{yang2022partial} introduced partial variable training (PVT) which involves freezing layers in clients and training a specific set of layers per client and aggregating them per layer for the server updates. 

\noindent\textbf{SER}
In order to deploy these models for real-time usage, we have to note that the availability of labeled data to the clients is extremely low, if not non-existent. Hence, semi-supervised FL frameworks are increasingly popular to train SER models with limited labeled data points per training client and a larger unlabelled set of data. The unlabelled sets of data are used in the supervised training by predicting pseudo labels. \citet{tsouvalas2022privacy} generated the pseudo labels for the unlabelled models and retains them based on the confidence measure of the label. Whereas \citet{feng2022semi} used multiview pseudo-labeling (\cite{xiong2021multiview}) coupled with uncertainty-aware pseudo-labeling selection process (\cite{rizve2021defense}) to generate the pseudo labels.

\begin{figure}[t]
    \begin{center}
        \includegraphics[width=\linewidth]{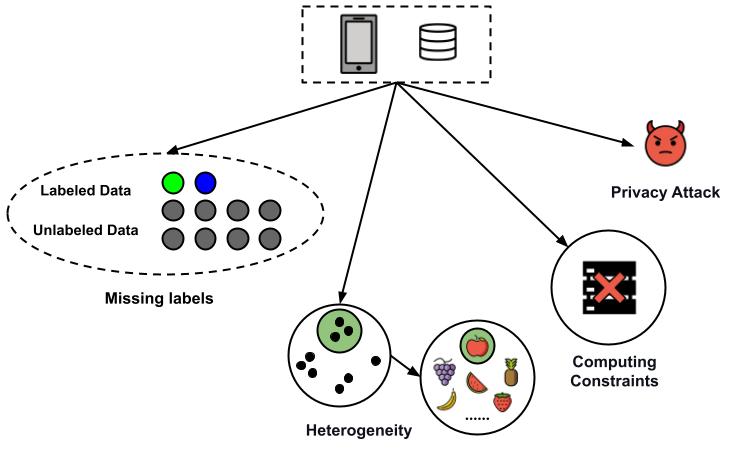}
    \end{center}
    \caption{Challenges in Federated speech-centric applications.}
    \label{fig:fed_challenges}
\end{figure}

\noindent\subsection{Challenges in FL}
Despite considerable improvements in privacy owing to the transition from a centralized to a decentralized training approach in FL, significant challenges remain for the ubiquitous adoption of FL models in speech processing. Although some speech processing tasks can attain the state of the art performance within parameter counts of 300k, tasks such as speech recognition and speech generation require about 100M parameters to obtain performance that is equivalent to state of art. However, clients in an FL are constrained by low computational power and model sizes. This provides an opportunity for future research into better optimization techniques or smaller models which can enable these tasks to be trained in a federated setting.

Another issue that arises while training models at clients in a supervised manner is the lack of clean labels. Therefore, semi-supervised and unsupervised techniques are of interest since they do not have to deal with the lack of clean labeled data. Some methods to tackle this issue in current works include employing a student-teacher framework providing weak labels on the local datasets and using external metadata to infer the labels based on user actions.

Moreover, despite user data not being transferred to the server, it has been shown that the gradients which are transferred are susceptible to a multitude of privacy attacks, such as data reconstruction attacks, property inference attacks, membership inference attacks, and poisoning attacks which adversely affect the privacy-preserving nature of FL. Although there have been numerous works on privacy attacks and defenses in federated learning, most of them focus on image or text tasks. Privacy attacks for speech processing tasks are a relatively under-explored area so are the defenses. However, it is an increasing field of interest owing to the large-scale adoption of voice assistants and their privacy concerns.

\noindent\subsection{Privacy Attacks in FL}
\citet{tomashenko2022privacy} proposed two attacks on ASR models in order to infer the speaker identity. One is purely statistical, and the other is a neural network-based attack to infer speaker identity based on the outputs of hidden layers from speaker models transmitted from the clients to the server. It has been demonstrated that this attack is successful with EER values obtained about 1-2\%. Similar property inference attacks were performed on SER models by \citet{feng2021attribute}, wherein the gender of the client was inferred from the model updates transmitted to the central server, which could either be the gradients or weight parameters. The attack is performed by shadow training (\cite{shokri2017membership}) on an open dataset similar to the target dataset, followed by using the model updates obtained during the shadow training to infer the gender of the speaker. A point to note is that in both the previous attacks, it is observed that the first hidden layer of the model provides the most information about the speaker identity. 

\noindent\subsection{Defences for FL}
Defenses for FL models mainly include the use of DP (\cite{dwork2008differential}), or Homomorphic Encryption (\cite{paillier1999public}). In the speech processing domain, \citet{feng22b_interspeech} deployed User-Level Differential Privacy (UDP) to mitigate property inference attacks from SER models. However, one limitation of the proposed defense is that when the adversary obtains multiple updates of the models, the performance of the defense degrades. \citet{chang2022robust} proposed a two-layered defense mechanism i.e., a combination of randomization and adversarial training in a federated setting to defend against FGSM, PGD, and Deepfool attacks for SER tasks.

\chapter{Bias and Fairness in Speech-centric Machine Learning}

\section{Fairness in Machine Learning}
The advance of machine learning technology has led to its ubiquitous application spanning several domains including healthcare, travel, and also the judicial system. While ML can alleviate human effort and promote automation, it has to be used cautiously to avoid potential biases to infiltrate the automatic decision-making process (\cite{van2022overcoming}). For example, a popular study investigated Recidivism Prediction Instruments (RPI) -- ML technology used to predict if a person who had committed a criminal offense in the past is likely to commit an offense in the future, and found that the popular COMPAS RPI was biased against black defendants (\cite{angwin2016machine}).

Similarly, evidence of bias has been found in face recognition (\cite{srinivas2019face,robinson2020face}), natural language applications~(\cite{lu2020gender}) as well as voice assistants (\cite{dheram2022toward}). As explained below, such bias can originate from either the training data, features/model or incorrect application of algorithm. Hence, mitigation methods have been proposed at each of these stages of the pipeline, from data curation to model deployment. In this section, we briefly delineate individual and group fairness, introduce different causes of bias, and finally delve into mitigation methods proposed in the speech domain. 

\noindent\subsection{Notions of Fairness:}

The fairness of machine learning algorithms is measured along one of two dimensions: 

\noindent a) \textit{Individual fairness:} tracks fairness/bias at the level of an individual member of a population, with the assumption being that similar individuals will be treated similarly (\cite{dwork2012fairness, kusner2017counterfactual}).

\noindent b) \textit{Group fairness:} measures relative bias between different subgroup populations of interest (\cite{hardt2016equality,berk2021fairness}). 
These sub-groups are typically divided along the lines of sensitive attributes of individuals such as gender, race, or age. 

\noindent\subsection{Causes of Bias:}
Unfair machine learning applications can be largely attributed to one of two main causes:

\noindent a) \textit{Data Bias}: 
Sources of data bias can be broadly categorized into measurement bias, omitted variable bias, representation bias, sampling bias and aggregation bias (\cite{mehrabi2021survey}).
These biases creep in either due to a) incorrect sampling of data from different subgroups, b) biased feature representations used for modeling, or c) misinterpretation of population statistics for subgroup statistics.
A common mechanism used to mitigate data bias due to imbalance in subgroups is to over-sample data from minority subgroups during training (\cite{dheram2022toward, fenu2020exploring}).

\noindent b) \textit{Algorithmic Bias}:
Algorithmic bias arises when an ML algorithm introduces bias in the system or amplifies existing bias in the data. 
Such bias permeates in the absence of carefully designed ML algorithms.
Algorithmic bias can be classified based on training data bias, bias in designing an algorithm or bias in deploying an algorithm (\cite{danks2017algorithmic}).
Training data bias can originate from an algorithm that propagates existing bias in the data (\cite{srinivas2019face}).
Bias in the design of an algorithm can be either a focus bias, wherein the algorithm uses features that are biased towards specific sub-groups, or processing bias, wherein the algorithm itself introduces bias as in the case of a statistically biased estimator.
Finally, bias can occur in the deployment or interpretation of an algorithm.
For example, using an algorithm outside the context it is developed for can lead to unfair results.
Similarly, using incorrect performance metrics that do not reflect the distribution of the data can lead to misinterpretation of the results.
Methods used to mitigate algorithmic bias include adversarial training and joint multi-task training (\cite{sun2018domain,tripathi2018adversarial,peri2022train}).

Different formulations have been proposed based on the fairness objective being targeted (\cite{verma2018fairness}). 
For example, some fairness metrics such as statistical parity and equal acceptance rate only consider the predicted outcome of a model.
More commonly used metrics, including equalized odds, predictive equality, predictive parity and equal opportunity further consider the true label in their fairness definition.

In speech processing applications, however, most work still use standard performance metrics to obtain fairness metrics (e.g., word error rate (WER) for ASR, equal error rate (EER) for ASV)).
As shown by \citet{peri2022train}, these metrics do not always hold, especially when applied to subgroups of population. 
In the following subsection, we outline different fairness related work including mitigation strategies in speech processing applications and present the different data resources that have been curated for fairness research. 

\begin{table}[]
\caption{Speech datasets for fairness evaluation}
\resizebox{\textwidth}{!}{
\begin{tabular}{llll}
\hline
\textbf{Dataset}   & \textbf{Task} & \textbf{\begin{tabular}[c]{@{}l@{}}Sensitive attributes \\ (\# classes)\end{tabular}} & \textbf{No. Hours} \\ \hline
Fairvoice (\cite{fenu2020exploring})   & ASV   & Gender (2), Age (9), Language (6) & 1700  \\ 
Casual Conversations (\cite{liu2022towards})  & ASR  & Gender (2), Skin Type (6)   & 572   \\    
TedTalk (\cite{acharyya2020fairyted})     & ASR           & Gender (3), Race (4)   & 564  \\ 
Artie Bias Corpus (\cite{meyer2020artie}) & ASR & Gender (3), Age (8), Accent (3) & 2.4 \\
Speech Accents (\cite{weinberger2011speech}) & ASR & Accents (7), Gender (2), Age    & <1 \\  \hline
\end{tabular}
}
\label{table:fairness_datasets}
\end{table}

\section{Fairness in Speech-centric Applications}

Compared with the flourishing of fairness research in other domains, such as facial analysis and natural language processing, the importance of addressing bias issues in speech processing has been underestimated for a long time. Up to today, there is still a limited amount of studies focusing on speech-related fairness. However, the accuracy degradation of a biased speech model on a specific demographic group not only leads to users' inconvenience but also makes the group believe the product is not designed for them. Therefore, it is essential to systematically evaluate the speech-related models' performance on fairness and investigate effective methods to mitigate the bias in speech models.
Several datasets have been proposed to advance fairness research in the speech domain (See Table \ref{table:fairness_datasets}. In addition to the task-specific labels (speaker/transcript for ASV/ASR), these datasets include labels for sensitive demographic attributes such as gender, race, age, accent, and language. 

\subsection{Fairness in Automatic Speaker Verification}

ASV systems suffer from bias in different demographic attributes, such as gender, age, language, and ethnic. For example, \citet{stoll2011finding} analyzed the performance difference of statistical speaker models regarding gender and age. However, the analysis is based on the verification scores while lacking deeper insights into the bias of the decision model and the distribution of speaker embeddings. As discussed earlier, these two are essential factors in clearly uncovering bias in ML models. 
More recently, \citet{peri2022train} investigated bias in ASV, showing the importance of distribution scores and also proposed fairness metrics for ASV instead of standard EER. They also explored mitigation strategies of adversarial training and multi-task training to reduce gender bias in ASV systems.

In addition to gender and age, ASV systems have also been found to be biased across languages. A study of UN-meetings (\cite{hebbar2021computational}) investigated the effect of language in speaker verification results and found that ASV degraded for Russian language as compared to English.
\citet{jin2022adv} forced the ASV learner to focus on poorly performing instances by weighting samples with an adversarial reweighting network and demonstrated that the reweighting method significantly improves the performance of ASV across different subgroups of gender and nationality.

NIST introduced new languages in their Speaker Recognition Evaluation protocol in 2016 (\cite{sadjadi20172016}) and 2018 (\cite{sadjadi20192018}), to investigate the influence of language in the speaker verification system.  In the 2021 VoxCeleb Speaker Recognition Challenge (VoxSRC) (\cite{brown2022vox}), the language attribute was added to the speaker verification track, aiming to encourage researchers to solve the performance degradation issue in the multi-lingual setting and boost the fairness in speaker verification. It is worth noting that the majority of the Fairness studies on the ASV task provide the evaluations using the VoxCeleb dataset series (\cite{nagrani2017voxceleb, chung2018voxceleb2}). 
\citet{chen2022exploring} explored the bias in speaker identification systems across different race groups and found that latinxs performed significantly worse than Caucasian speakers.
Recently, \citet{fenu2020exploring} explored the fairness in deep learning-based ASV systems by collecting a speaker dataset -- Fairvoice, conducting performance analysis on EER and score (FAR, FRR) distributions, and providing more understanding of how diverse speaker verification is correlated with demographic attributes.  
To standardize the ASV fairness evaluation, \citet{tous2021sveva} devised a framework to provide comprehensive fairness evaluation metrics and visualization methods to present model's fairness across subgroups.

\subsection{Fairness in Automatic Speech Recognition}

In the past decade, due to the development of deep learning and the availability of large-scale speech and language databases, the word error rate (WER) of ASR models has decreased to a satisfactory level in many languages.  However, the fairness of ASR systems still raises the interests of researchers from psychological, sociology, and engineering backgrounds (\cite{mengesha2021impact, rajan2022aequevox}). \citet{mengesha2021impact} investigated the ASR failure on African American Vernacular English and demonstrated the detrimental impact on African American users from the psychological perspective.  In addition to proposing a set of methodologies to model the users' feelings and experience in fairness research, Mengesha \textit{et al.} also encourages researchers to spot more light on fairness AI. \citet{rajan2022aequevox} introduced an automated testing framework (AEQUEVOX) for evaluating the fairness of ASR systems. By conducting extensive fairness experiments on four datasets with three commercial ASRs, \citet{rajan2022aequevox} validated the ASR fairness violation on non-native English, female, and Nigerian English speakers.
With the recent advent of self-supervised learning, approaches such as wav2vec2 (\cite{baevski2020wav2vec}) are being widely used in a variety of speech recognition related tasks.  \citet{boito2022gender} investigated the impact of pretrained data distribution on the fairness performance across subgroups.  By pretraining the wav2vec 2.0 with gender-specific and different proportion of gender data, it is demonstrated that the fairness is related to downstream integration and balanced-gender pretraining data does not necessarily reduce bias.

To mitigate the bias in ASR on the groups across geographic locations and demographic attributes, \citet{dheram2022toward} proposed an initial method, oversampling under minority groups and undersampling majority groups, to reduce the performance gap between different cohorts. \citet{liu2022towards} presented results from multiple ASR models on the Casual Conversations dataset and observed the significant WER difference across gender and ethic.  To accurately evaluate the ASR fairness issue on racial demographics, \citet{liu2022model} adopted mixed-effects Poisson regression to mitigate the negative influence from nuisance factors, such as speaker, context, phoneme, prosody, etc.

\subsection{Fairness in Speech Emotion Recognition}

Fairness in SER is important across multiple areas because the performance differences resulting from gender and race are significant in most of emotion recognition scenarios (\cite{sharma2021survey}). \citet{gorrostieta2019gender} investigated gender-based bias in speech emotion recognition and mitigated unwanted bias through adversarial training and additional weight for the objective function. In recent years, the large-scale self-supervised learning (SSL) model has become popular in computer vision, language processing, and audio processing. With the widespread use of SSL, \citet{wagner2022dawn} have shone light on the performance variance and demonstrated that transformer-based SSL models have moderate fairness scores in the SER area.  

\chapter{Future Directions}

In this section, we discuss the main challenges and potential research opportunities for speech-centric trustworthy machine learning in order to inspire readers to research this field further. 

\section{Privacy}

In recent years, self-supervised learning speech models such as Wav2Vec 2.0 (\cite{baevski2020wav2vec}) and Whisper (\cite{radford2022robust}) have established the SOTA performance for many downstream speech tasks such as ASR. However, privacy-related topics, such as MIAs, on these emerging model techniques have not been explored extensively. As we discussed earlier, \citet{tseng22_interspeech} was the only one to investigate MIA threats to SSL speech models. Their results imply that the latent speech representation of SSL models holds the membership information of the input speech signals. However, they only perform some preliminary defenses on MIAs and findings from these defenses are limited. Therefore, it is valuable and critical to extend the work presented in \citet{tseng22_interspeech} to broader SSL speech models, downstream speech tasks, and MIA mitigation strategies.

In addition, there exist more property inferences where current PIAs have not been explored but are of demand in speech-centric ML, e.g., age, health status, etc. As our review points out, the majority of literature focuses on gender obfuscation, while none of the studies attempts to generalize the existing approaches to demographics like age or race. Apart from the lack of exploration of broader demographics, many studies only evaluate their works on several clean-audio benchmarks like Librispeech (\cite{panayotov2015librispeech}) and IEMOCAP (\cite{busso2008iemocap}), while the robustness and efficacy of many proposed privacy-enhancing approaches on the more dynamic recording conditions are unknown.

As we also presented in the Privacy review section, Federated Learning has become an emerging research topic in almost every field of ML. Nevertheless, compared to NLP and CV domains, the FL on speech-centric tasks stays largely unexplored. One of the significant challenges is to enable the training of the ASR system in the FL setting. Due to the nature of the ASR modeling, expensive computations are typically mandatory, while most mobile devices cannot afford to train and run these heavy computing models. Therefore, it is urgent and essential to investigate efficient FL training approaches for ASR models. Last but not least, most literature to date focuses on dealing with missing labels and decoupling heterogeneity conditions in FL, and fewer works are targeting privacy risks in federated speech learning. As a result, an exciting but critical research direction is systematically studying the privacy risks in federated speech modeling.

\section{Safety}

While consideration has been given to speech-specific evasion attacks, there has been far less work on audio-modality defenses. Speech signals possess unique structural and temporal properties that distinguish them from images. Traditional defense methods from the computer vision domain fail to fully leverage these attributes in aid of adversarial robustness. Furthermore, there have been some works such as by \citet{qin2019imperceptible} and \citet{yakura2018robust} that consider `over the air attacks' in which the adversarial audio travels to the model by means of an acoustic channel (as opposed to being directly fed to the model input). More work is needed in this area to gauge the feasibility and threat level of evasion attacks in this more realistic setting. Finally, there is relatively little work on poisoning attacks and defenses for speech systems. Given that large-scale datasets are increasingly procured from unverified sources (i.e. internet posts, client data in FL), it is critical that we better understand the potential risks and how to mitigate them.

\section{Fairness}
As outlined in the previous section, research on fairness and bias mitigation in the speech domain is limited to a few preliminary works. 
These are typically restricted to fairness evaluations on small attribute-balanced datasets for either gender or accent. 
With the introduction of newer datasets (Table \ref{table:fairness_datasets}), we hope to see more fairness studies along the demographic attributes of ethnicity, language, etc. as well as intersectional attributes (e.g, female Spanish vs male Spanish).
Furthermore, there has been little work exploring the extent of bias in recent SOTA models such as Wav2Vec2.0 (\cite{baevski2020wav2vec}) and Whisper (\cite{radford2022robust}) for both ASR and ASV. 
There has also been little to no work in other speech tasks such as SER.
Finally, the need for fairness-specific metrics is highlighted by \citet{peri2022train}, with most of the existing literature using standard metrics such as EER and WER. 

\section{Balance between Fairness, Privacy, and Safety}

Despite the tremendous effort in designing trustworthy machine learning techniques over the last decade, most of these applications attempt to isolate one single aspect of trustworthiness in their modeling work. Consequently, a limited amount of work explores the impact of mitigating one trustworthy risk over other trustworthy challenges. Recently, \citet{chang2021privacy} demonstrated that privacy and fairness are often opposed to each other in trustworthy machine learning. The author showed that increasing model fairness requires optimizing objectives that typically constrain the model to perform equally on every subgroup, leading the model to memorize training samples from the unprivileged subgroups. Hence, enhancing model fairness frequently raises significant privacy concerns in exposing the private information of the training data.

Apart from investigating the relationship between fairness and privacy, \citet{xu2021robust} found that safety-awareness learning poses a disparate impact on the fairness risk of subgroups. The author also proposed a Fair-Robust-Learning (FRL) framework to enhance the model fairness while performing adversarial defenses. However, both of these works solely investigate computer vision applications, while the interplay between privacy, fairness, and safety in speech-centric applications remains largely unexplored. It is, therefore, critical for future researchers to investigate the interactions between different trustworthiness aspects in speech-centric applications.

\section{Trustworthy Multimodal Applications}

Besides speech-centric applications that we discussed in this article, speech signals have broader usage cases in diverse multimodal applications, such as multimedia applications, audio-visual understanding, and multimodal sentiment analysis (\cite{jiang2021review}). Typically, multimodal applications are reported with a higher system performance than unimodal models. Consequently, one promising future research direction is to explore speech-based multimodal applications that attempt to achieve more satisfactory balances between fairness, privacy, and safety.

\chapter{Conclusion}

As machine learning becomes more prevalent in speech-centric modeling, the scientific community has also become more aware of concerns and risks over the trustworthiness of these systems. This survey paper aims to provide a comprehensive and systematic summary of the recent efforts made to protect privacy, ensure fairness, and defend against adversarial attacks in these speech-centric systems. Increasingly, we identify several open problems of importance to address, such as investigating the extent of bias in recent SOTA models. Through this review, we hope to provide the necessary knowledge for future research in speech-centric trustworthy ML.

\chapter{Acknowledgement}
This work was supported by DARPA (Grant No: HR00112020009) and USC Amazon Center for Secure and Trusted Machine Learning.

\backmatter  

\printbibliography

\end{document}